\pdfoutput=1
\documentclass[fleqn,usenatbib]{mnras}

\usepackage{mathptmx}
\usepackage{txfonts}
\usepackage{type1ec}
\usepackage[pdftex]{graphicx}

\usepackage[T1]{fontenc}
\usepackage{ae,aecompl}

\usepackage{graphicx}	% Including figure files
\usepackage{amssymb}	% Extra maths symbols
\usepackage[usenames,dvipsnames]{xcolor}
\newcommand{\mdota}{\dot{M}_{\rm acc}}
\newcommand{\mdotw}{\dot{M}_{\rm wind}}
\newcommand{\mdotin}{\dot{M}_{\rm in}}
\newcommand{\mdotout}{\dot{M}_{\rm out}}
\newcommand{\aO}{a_{\rm 0}}
\newcommand{\msun}{M_{\rm \odot}}
\newcommand{\mstar}{M_{\rm \star}}

\newcommand{\vk}{v_{\rm K}}
\newcommand{\vr}{v_{\rm r}}
\newcommand{\vrO}{v_{\rm r0}}
\newcommand{\vphi}{v_{\rm \phi}}
\newcommand{\vz}{v_{\rm z}}
\newcommand{\vA}{v_{\rm A}}
\newcommand{\vAO}{v_{\rm A0}}
\newcommand{\hthermal}{h_{\rm T}}
\newcommand{\cs}{c_{\rm s}}
\newcommand{\zh}{z_{\rm h}}
\newcommand{\zb}{z_{\rm b}}
\newcommand{\zs}{z_{\rm s}}
\newcommand{\rhoO}{\rho_{\rm 0}}

\newcommand{\rhob}{\rho_{\rm b}}

\newcommand{\rhosnorm}{\tilde{\rho}_{\rm s}}
\newcommand{\ElsasserO}{\Lambda_{\rm 0}}
\newcommand{\etaA}{\eta_{\rm A}}
\newcommand{\etaH}{\eta_{\rm H}}
\newcommand{\etaO}{\eta_{\rm O}}
\newcommand{\etaOnorm}{\tilde{\eta}_{\rm O}}
\newcommand{\etaHnorm}{\tilde{\eta}_{\rm H}}
\newcommand{\etaAnorm}{\tilde{\eta}_{\rm A}}
\newcommand{\SigmaO}{\Sigma_{\rm 0}}

\newcommand{\rin}{r_{\rm in}}
\newcommand{\rout}{r_{\rm out}}
\newcommand{\epsilonB}{\epsilon_{\rm B}}

\title[Winds from protostellar accretion discs]{Centrifugally driven winds from protostellar accretion discs. I -- Formulation and initial results.}

\author[C. A. Nolan]{C. A. Nolan$^{1}$\thanks{E-mail:
chris.nolan@anu.edu.au (CAN); raquel.salmeron@anu.edu.au (RS); christoph.federrath@anu.edu.au (CF); geoff.bicknell@anu.edu.au (GVB); ralph.sutherland@anu.edu.au (RSS)}, R. Salmeron$^{1,2}$\footnotemark[1]\thanks{Present address: Airservices Australia, 25 Constitution Ave., Canberra, ACT 2601, Australia}, C. Federrath$^{1}$\footnotemark[1], G. V. Bicknell$^{1}$\footnotemark[1] and
\newauthor R. S. Sutherland$^{1}$\footnotemark[1]
\\
$^{1}$Research School of Astronomy and Astrophysics, Australian National University, Canberra, ACT 2611, Australia \\
$^{2}$Mathematical Sciences Institute, Australian National University, Canberra, ACT 2601, Australia}

\date{Accepted XXX. Received YYY; in original form ZZZ}

\pubyear{2016}

\begin{document}
\label{firstpage}
\pagerange{\pageref{firstpage}--\pageref{lastpage}}
\maketitle

\begin{abstract}
Protostellar discs play an important role in star formation, acting as the primary mass reservoir for accretion onto young stars and regulating the extent to which angular momentum and gas is released back into stellar nurseries through the launching of powerful disc winds. In this study, we explore how disc structure relates to the properties of the wind-launching region, mapping out the regions of protostellar discs where wind launching could be viable. 
We combine a series of 1.5D semi-analytic, steady-state, vertical disc-wind solutions into a radially extended 1+1.5D model, incorporating all three diffusion mechanisms (Ohm, Hall and ambipolar). 
We observe that the majority of mass outflow via disc winds occurs over a radial width of a fraction of an astronomical unit, with outflow rates attenuating rapidly on either side. We also find that the mass accretion rate, magnetic field strength and surface density profile each have significant effects on both the location of the wind-launching region and the ejection/accretion ratio $\mdotout/\mdotin$. Increasing either the accretion rate or the magnetic field strength corresponds to a shift of the wind-launching region to smaller radii and a decrease in $\mdotout/\mdotin$, while increasing the surface density corresponds to launching regions at larger radii with increased $\mdotout/\mdotin$. Finally, we discover a class of disc winds containing an ineffective launching configuration at intermediate radii, leading to two radially separated regions of wind launching and diminished $\mdotout/\mdotin$. We find that the wind locations and ejection/accretion ratio are consistent with current observational and theoretical estimates.
\end{abstract}


%%%%%%%%%%%%%%%%%%%%%%%%%%%%%%%%%%%%%%%%%%%%%%%%%%

%%%%%%%%%%%%%%%%% BODY OF PAPER %%%%%%%%%%%%%%%%%%

%----------------------------------------------
% INTRODUCTION
%----------------------------------------------
\section{Introduction} \label{sec:introduction}

Protostellar discs are an integral part of the star and planet formation processes \citep{Li:2014tx}. They are formed via angular momentum conservation as a pre-stellar core collapses, and become the primary source of material for the young, central protostar as it builds up to its final mass. For material to accrete through the disc and onto the growing star, angular momentum, since conserved, must be redistributed within the disc or ejected from the system \citep{Turner:2014tw}. This process may be facilitated by bipolar outflows and winds, which are also important for star formation, including the initial mass function and star formation rate \citep{Federrath:2014ii, Frank:2014wl, Krumholz:2014wk, Li:2014tx, Offner:2014vw, Padoan:2014vb, Federrath:2015fw}.

Observations show that the occurrence of accretion and outflow are correlated \citep[e.g.][]{Bally:2007uy}, and this correlation is marked by accretion diagnostics and outflow signatures. The rates of accretion and outflow in these systems are also correlated, with bipolar jets expelling between 0.1--0.2 times the amount of material accreted onto the star \citep{Cabrit:2007hz, Ellerbroek:2013bf, Watson:2015uo}. Any theory of wind-launching must reproduce these correlations in order to be viable.

Mass transport through protostellar discs onto their host stars is likely to take place via a combination of different accretion mechanisms operating within different regions of the disc. Two main accretion mechanisms stand out: magnetohydrodynamic (MHD) turbulence induced by the magnetorotational instability \citep[MRI,][]{Balbus:1991fi, Balbus:1998du, Bai:2014ih}, and centrifugally driven winds (CDWs) \citep[][hereafter BP82]{Blandford:1982vy}. In recent years, through developments in modelling of turbulent diffusive discs, there has been serious doubts raised as to the effectiveness of MRI turbulence in the range of 1--10 astronomical units (au) \citep[see the recent review of][]{Turner:2014tw}. The MRI is sensitive to the dominant magnetic diffusion mechanism in the disc and within this range Hall diffusion is expected to be the strongest at the disc midplane \citep{Wardle:2012cs}. \citet{Kunz:2013jp} found that if the Hall term is large enough, the MRI turbulence transforms into self-sustaining zonal structures with poor angular momentum transport. Similarly, \citet{Wardle:2012cs} discovered that if the magnetic field is anti-parallel to the rotation axis and small dust grains are present, the vertical column available to the MRI is insignificant. This has led to a renewed focus on disc winds as the dominant angular momentum removal mechanism in this region.

%Literature review
Since the pioneering work of BP82 -- which focussed on radio jets launched from active galactic nuclei -- and its re-application to protostellar jets by \citet{Pudritz:1983wn,Pudritz:1986uu}, there have been many advances in our understanding of disc winds and the methods used to model them. Early work established the connection between disc properties, mass loading of winds and angular momentum transport by including the disc as a specific region within a 2D simulation domain \citep[e.g.][]{Shibata:1985vb,Shibata:1986to}, or as a fixed boundary condition \citep[e.g.][]{Ustyugova:1995jp,Ouyed:1997wg}. Subsequent studies involved self-similar solutions \citep{Li:1995bt, Li:1996gm, Ferreira:1997wp}, which emphasised the importance of the magnetic field in driving the wind, and in some cases were able to predict the large-scale behaviour of the flow \citep{Teitler:2011vj}. More recently, shearing-box simulations have been employed to investigate the properties of winds using realistic microphysics \citep{Suzuki:2009wl, Suzuki:2010ul, Fromang:2013vf, Bai:2013ue, Bai:2013bx, Lesur:2013ha, Simon:2013fw, Bai:2014ea}. These simulations reinforce the role disc winds play in mass transport in protostellar discs and accretion onto the protostar. 

%Local simulations are inadequate
Local simulations such as the ones listed above are important for understanding the dynamical behaviour in particular parts of the disc, but are limited in scope. The challenge for global disc models is to accurately represent non-ideal effects while modelling large portions of the disc. This task was first attempted by \citet{Dzyurkevich:2010kp} in the context of the MRI, using a fixed Ohmic resistivity distribution, and more recently \citet{Gressel:2015ks} added ambipolar diffusion and time-dependent gas-phase electron and ion fractions.

%Gap in the literature
In this paper, we develop a new approach for investigating protostellar disc winds, which allows for the resolving of steady-state axisymmetric wind solutions in the non-ideal MHD regime. We achieve this by linking together a set of radially localized, vertical 1.5D solutions of the type designed by \citet[][hereafter WK93]{Wardle:1993dm}, \citet[][hereafter KSW10]{Konigl:2010it} and \citet[][hereafter SKW11]{Salmeron:2011bd}, to create a radially extended, 1+1.5D model of the wind-launching region in axisymmetric cylindrical coordinates ($r$, $z$). We employ these models to investigate how the structure of the underlying disc affects the properties of the wind. We find that the location of the wind-launching region and the ejection/accretion ratio $\mdotout/\mdotin$ are significantly influenced by the mass accretion rate, magnetic field strength and surface density profile, while still satisfying observational and theoretical constraints. Most importantly, we find that the wind-launching region is radially localized, with the outflow rate decreasing rapidly at larger and smaller radii, and we discover a class of disc winds containing an ineffective launching configuration at intermediate radii.

%Summarize the paper structure
We begin by summarizing the numerical method in Section \ref{sec:method}, before providing a comprehensive description of the 1.5D and 1+1.5D models in Sections \ref{sec:1.5D} and \ref{sec:2d}, respectively. In Section \ref{sec:comp}, we present a detailed analysis of two distinct wind models, followed by a more general parameter study in Section \ref{sec:change_var}. We discuss the implications of our results in Section \ref{sec:discussion} and summarize our conclusions in Section \ref{sec:conclusions}.

%----------------------------------------------
% METHODOLOGY INTRODUCTION
%----------------------------------------------
\section{Method summary} \label{sec:method}

\begin{figure}
	\centering
	\includegraphics[width=84mm]{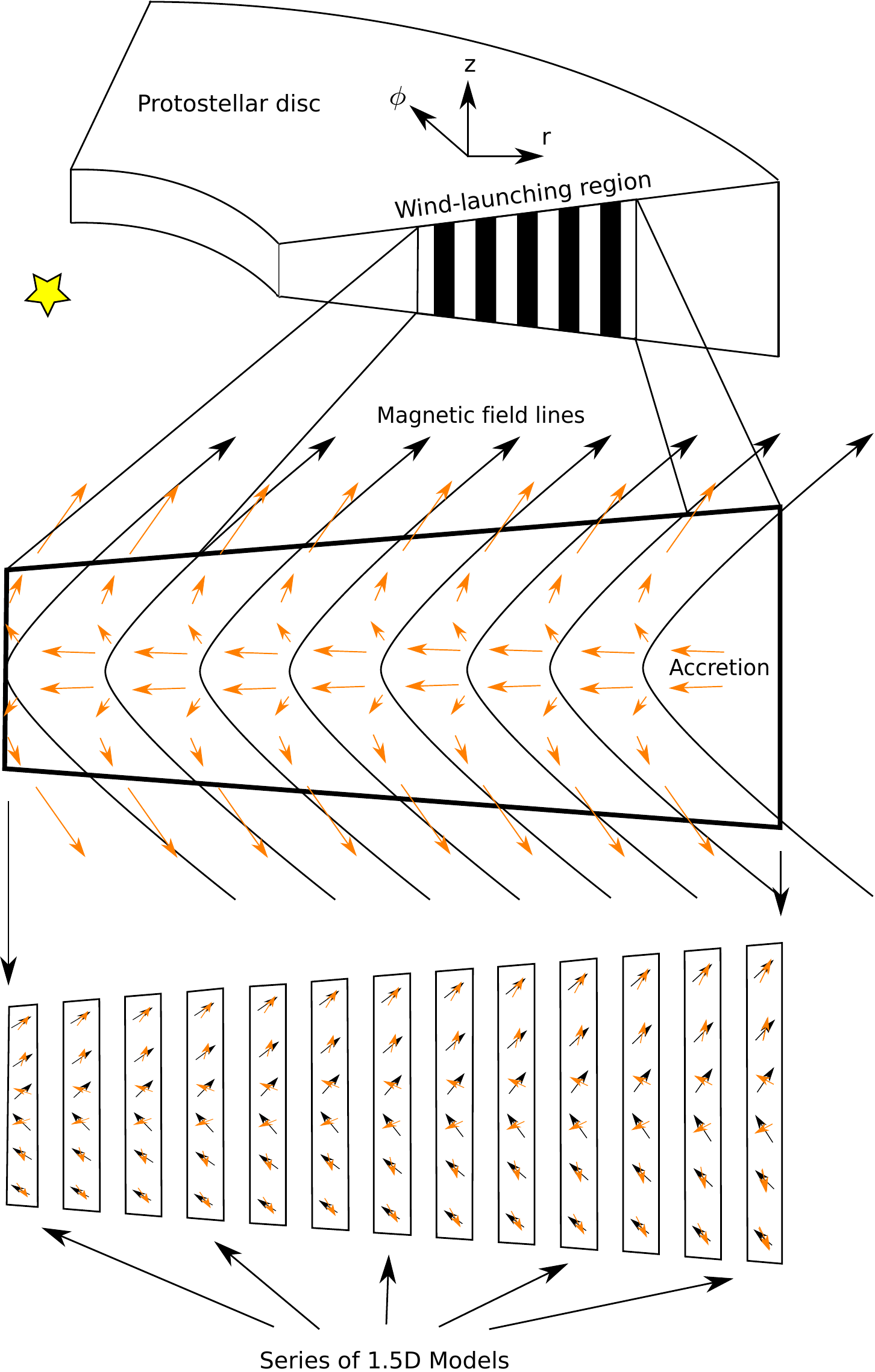}
	\caption{Schematic diagram showing the construction of the 1+1.5D models. Each model consists of a series of vertical, 1.5D radially-localized solutions positioned at consecutive radii, from the inner radius of the wind-launching region $\rin$ to its outer radius $\rout$. All 1.5D solutions provide disc/wind properties from the disc midplane up to the sonic point $\zs$, including radial, azimuthal and vertical vector components.}
	\label{fig:schematic}
\end{figure}

In the following sections, we describe our method for modelling the wind-launching region of protostellar discs. We provide here a brief summary of the basic method before moving on to the full derivations in Sections \ref{sec:1.5D} and \ref{sec:2d}. 

To model the wind region, we link together a number of vertical, axisymmetric, 1.5D radially-localized solutions in radius, in such a way as to form a self-consistent 1+1.5D model in (r,z)-coordinates (see Fig. \ref{fig:schematic}). Each 1.5D model is the integrated solution to a set of six ordinary differential equations (ODEs) in the vertical coordinate $z$, from the disc midplane up to the sonic point $\zs$\footnote{Ideal MHD flows have three critical surfaces, beginning with the slow magnetosonic point. However, when the flow is diffusive, the thermal sonic point becomes the first critical point of the flow \citep[see][pp. 326]{Konigl:2011tk}} (see Section \ref{sec:1.5D:build}), and provides normalized values for the density $\rho$, velocity $\mathbf{v}$, magnetic and electric fields ($\mathbf{B}$ and $\mathbf{E}$ respectively) and current density $\mathbf{J}$ within this vertical range. 

Each 1.5D solution is uniquely characterized by a number of dimensionless parameters listed in Section \ref{sec:1.5D:parameters}. By specifying conditions at the disc midplane (e.g. $\rhoO$, $B_{\rm 0}$, and the temperature $T_{\rm 0}$) at a particular radius, and calculating the diffusion coefficients $\eta$ using our ionization model (see Appendix \ref{sec:ionization}) we can calculate these input parameters and solve for the vertical structure of the disc and wind. However, instead of specifying all of the conditions at the disc midplane directly, we obtain them by prescribing the values of the surface density ($\Sigma$), the local mass accretion rate ($\mdota$), the midplane ratio of the Alfv\'{e}n speed to the isothermal sound speed ($\aO$) and the vertically isothermal temperature ($T_{\rm 0}$) at that radius, in addition to the stellar mass ($\mstar$), and iterate on the midplane density ($\rhoO$) and radial velocity ($\vrO$) until the vertically integrated $\Sigma$ and $\mdota$ of the solution matches the prescribed $\Sigma$ and $\mdota$. Hence, by assigning values for $\Sigma$, $\mdota$, $\aO$ and $T_{\rm 0}$ at a particular radius within the disc, orbiting around a protostar of mass $\mstar$, we can calculate the vertical structure of the disc and wind at that point.

In order to create a 1+1.5D model, we assign the disc parameters $\Sigma$, $\mdota$, $\aO$ and $T_{\rm 0}$ at each radius, which allows us to calculate the vertical structure of the disc across a range of radii. However, these parameters must be chosen so that the 1+1.5D model does not violate the equations of non-ideal MHD, in the limit of a disc that is in a steady state and is geometrically thin, vertically isothermal, nearly Keplerian and in dynamical equilibrium in the gravitational potential of the central protostar (KSW10). To do this, we tie each pair of radially adjacent solutions together using mass conservation, and check to make sure that the thin disc approximation is not violated. We also ensure that the \mbox{$\nabla \cdot \mathbf{B} = 0$} condition is satisfied (see Section \ref{sec:2d:const} for more details). 

Finally, the inner and outer edges of our 1+1.5D model are defined by four constraints (listed in Section \ref{sec:1.5D:const}), which set the region of parameter space in which the 1.5D solutions are physically viable (WK93, KSW10). Hence we arrive at a self-consistent 1+1.5D model, which maps out the structure of the wind-launching region of a protostellar disc from the midplane up to the sonic point. We note here that these models are still localized in that they do not take into account the global magnetic field structure (being limited in height by the thin disc approximation) and are also simplified by the use of parametrized conductivity profiles.

%----------------------------------------------
% RADIALLY-LOCALIZED WIND-DRIVING DISC MODELS
%---------------------------------------------- 
\section{1.5D radially-localized disc wind models} \label{sec:1.5D}

The 1.5D models of KSW10 form the basis of our 1+1.5D approach to disc winds. Hence, before describing our 1+1.5D models in Section \ref{sec:2d}, we pause here to summarize the properties and derivation of the 1.5D solutions.

Each 1.5D solution assumes that the disc is in a steady state, geometrically thin, vertically isothermal, nearly Keplerian, and is in dynamic equilibrium within the gravitational potential of the central protostar. Within the disc, the degree of ionization is low due to radiation shielding, causing non-ideal effects to become important. We incorporate finite conductivity effects via a conductivity tensor formulation \citep[e.g.][]{Wardle:1999dy}, allowing the three basic field-matter diffusion mechanisms (Ohm, Hall and ambipolar) to be included without the need for separate equations for each fluid component (KSW10). For reference, the Ohm, Hall and ambipolar diffusion coefficients in the limit of a weakly-ionized ion-electron plasma are
\begin{equation}
\etaO = \frac{c^2 m_{\rm e} \gamma_{\rm e} \rho}{4 \pi e^2 n_{\rm e}}, \label{eqn:ohmdiff_ext}
\end{equation}
\begin{equation}
\etaH = \frac{cB}{4 \pi e n_{\rm e}}, \label{eqn:halldiff_ext}
\end{equation}
and
\begin{equation}
\etaA = \frac{B^2}{4 \pi m_{\rm i} \gamma_{\rm i} \rho n_{\rm i}} \label{eqn:ambidiff_ext}
\end{equation}
respectively \citep[e.g.][]{Wardle:2012cs}, where $c$ is the speed of light, $e$ is the charge of an electron, and for the electron and ion subscripts `e' and `i' respectively, we have the particle mass $m_{\rm e, i}$, the number density $n_{\rm e, i}$ and $\gamma_{\rm e, i} = \langle \sigma v \rangle_{\rm e, i} /(m_{\rm e, i} + m)$, where $\langle \sigma v \rangle_{\rm e, i}$ is the rate coefficient for collisional momentum transfer between the charged species and the neutrals.

In the remainder of this section we define the model parameters that characterize each radially-localized solution, summarize the method of solving for the vertical structure of the disc using these parameters as boundary conditions, and define the requirements for these solutions to be physically viable, before progressing on to the development of the 1+1.5D framework in Section \ref{sec:2d}.

%---------- Parameters ----------
\subsection{Parameters} \label{sec:1.5D:parameters}

Each 1.5D radially-localized disc-wind solution is described by six dimensionless parameters, which determine the boundary conditions of the 1.5D problem (see Section \ref{sec:1.5D:build}). They are:
\begin{description}
\item[(i)] The ratio of the Alfv\'{e}n speed ($\vAO$) to the isothermal sound speed ($\cs$) at the disc midplane
\begin{equation}
\label{a0_def}
\aO \equiv \frac{\vAO}{\cs} = \frac{B_{\rm 0}}{\sqrt{4 \pi \rhoO}} \frac{1}{\cs}. \label{eqn:a0}
\end{equation} 
This parameter quantifies the strength of the ordered magnetic field that threads the disc.
\item[(ii)] The ratio of the gravitational tidal scale height $\hthermal$ to the disc radius 
\begin{equation}
\frac{\hthermal}{r} = \frac{\cs}{\vk}, \label{eqn:hthermal}
\end{equation}
where $\vk$ is the Keplerian velocity. This parameter provides a measure of the geometric thickness of the disc and also constrains physically viable solutions (see equation \ref{eqn:const4}). It was used by SKW11 to match solutions to BP82-type winds.
\item[(iii)] The midplane ratios of the magnetic diffusivity components\footnote{Previously, KSW10 used the midplane conductivity ratios $[\sigma_{\rm P}/\sigma_{\rm \perp}]_{\rm 0}$ and $[\sigma_{\rm \perp}/\sigma_{\rm O}]_{\rm 0}$ to characterize solutions (where \mbox{$\sigma_{\rm \perp} = \sqrt{\sigma_{\rm H}^2 + \sigma_{\rm P}^2}$} and the subscripts O, H and P denote the Ohm, Hall and Pedersen conductivities respectively), however in this paper we adopt the diffusivity ratios as they are more intuitively connected to the three diffusivity regimes.}, 
\begin{equation}
\left[\frac{\etaH}{\etaO}\right]_{\rm 0} \qquad \mbox{ and} \qquad \left[\frac{\etaA}{\etaO}\right]_{\rm 0}. \label{eqn:diffratios}
\end{equation}
In general, the diffusivity components and their ratios vary with height above the disc midplane, reflecting the change in disc conditions with height. However we adopt the simplification of KSW10 in which the ratios described by equation (\ref{eqn:diffratios}) are constant with height, $z$. Specifically, we scale the components of the conductivity tensor with the density and the magnetic field strength as $\rho/B^2$, so that the field-matter coupling (see below) is constant with height. This simplification will be relaxed in a subsequent paper.
\item[(iv)] The midplane Elsasser number
\begin{equation}
\ElsasserO = \left( \etaHnorm^2 + \left( \etaAnorm + \etaOnorm \right)^2 \right)^{-1/2}, \label{eqn:elsasser}
\end{equation}
where
\begin{equation}
\mathbf{\tilde{\eta}} = \frac{\mathbf{\eta}}{\vAO^2 / \Omega_{\rm K} }, \label{eqn:normalizations}
\end{equation}
and $\Omega_{\rm K}$ is the Keplerian angular velocity. The Elsasser number measures the degree of coupling between the magnetic field and the neutrals, with the regimes of weak and strong coupling prescribed by $\ElsasserO \ll 1$ and $\ElsasserO \gg 1$ respectively. For future reference, the Elsasser numbers describing each diffusivity regime are defined as follows:
\begin{equation}
\Lambda_{\rm O} = \frac{1}{\etaOnorm} \mbox{,} \qquad \Lambda_{\rm H} = \frac{1}{\etaHnorm} \mbox{,} \qquad \Lambda_{\rm A} = \frac{1}{\etaAnorm}. \label{eqn:elsasser3}
\end{equation}
\item[(v)] The inward radial Mach number at the midplane
\begin{equation}
\epsilon \equiv \frac{-v_{\rm r0}}{\cs}, \label{eqn:epsilon}
\end{equation} 
which is a free parameter of the disc solution. We determine its value for the 1+1.5D model by constraining the local accretion rate $\mdota$ at each radius (see Section \ref{sec:2d:build}).
\item[(vi)] The normalized azimuthal component of the electric field $\mathbf{E}$ 
\begin{equation}
\epsilonB \equiv \frac{-cE_{\rm \phi0}}{\cs B_{\rm z}}, \label{eqn:epsilon_b}
\end{equation}
which measures the radial drift of the poloidal magnetic field lines. WK93, using a similar radially localized model, derived solutions for positive and negative values of $\epsilonB$ and found that configurations with the same value of ($\epsilon - \epsilonB$) were similar. This suggests that setting $\epsilonB = 0$ should not significantly impact the generality of the results (see Section \ref{sec:discussion} and Appendix A of KSW10). We adopt $\epsilonB = 0$ for the remainder of the paper.
\end{description}

%---------- Numerical integration of the localized disc equations ----------
\subsection{Numerical integration of the localized disc equations} \label{sec:1.5D:build}

The dimensionless parameters listed in Section \ref{sec:1.5D:parameters} are used to derive the boundary conditions for solving a set of six ODEs in the normalized vertical coordinate $\tilde{z} = z/\hthermal$, where the disc scale height $\hthermal$ is defined by equation (\ref{eqn:hthermal}). These ODEs are derived from the equations of non-ideal MHD using the thin disc approximation, and determine the vertical structure of $\rho$, $v_{\rm r}$, $v_{\rm \phi}$, $B_{\rm r}$, $B_{\rm \phi}$ and $E_{\rm r}$ in dimensionless form. What follows is a brief summary of the method of vertical integration for these equations; we refer the reader to SKW11 for a more comprehensive description.

To solve for the vertical structure of the disc, we begin by assigning the midplane values of $\rho$, $v_{\rm r}$, $v_{\rm \phi}$, $B_{\rm r}$, $B_{\rm \phi}$ and $E_{\rm r}$ in terms of the dimensionless parameters listed in Section \ref{sec:1.5D:parameters} (see equations 18 and $20-22$ of SKW11). We then guess the midplane value of $\tilde{v}_{\rm z0} = v_{\rm z0}/\cs$ and the position of the sonic point $\tilde{z}_{\rm s}$, and integrate from the midplane towards $\tilde{z}_{\rm s}$. If the guessed value for $\tilde{v}_{\rm z0}$ is too high then $\tilde{v}_{\rm z}$ diverges, and if it is too low, $\tilde{v}_{\rm z}$ peaks and begins to decrease with $\tilde{z}$, which is unphysical. This gives upper and lower limits for the value of $\tilde{v}_{\rm z0}$. We then use an iterative bisection method to improve upon $\tilde{v}_{\rm z0}$ until we are close enough to the physical solution ($\tilde{v}_{\rm zs} = 1$) to estimate the position of $\tilde{z}_{\rm s}$ and the values of the variables there. We then simultaneously integrate from $\tilde{z} = \tilde{z}_{\rm s}$ and the midplane ($\tilde{z} = 0$) to an intermediate fitting point (usually $\sim 0.7$--0.9 $\tilde{z}_{\rm s}$), and adjust the guessed variables at each end iteratively until the solution converges.

%---------- Constraints for physically viable solutions ----------
\subsection{Constraints for physically viable solutions} \label{sec:1.5D:const}

As previously shown by WK93 and KSW10, viable wind-driving disc solutions for which $\ElsasserO$ is not $\ll 1$ (where $\ElsasserO \ll 1$ is indicative of very weak field-matter coupling) exist within a limited region of parameter space. This region is determined by the following four requirements:

\begin{description}

\item[(i)] \textit{Sub-Keplerian flow}: The flow remains sub-Keplerian ($\vphi < \vk$) within the disc. Super-Keplerian flow below the disc surface would require that the excess rotation be balanced by inwardly directed radial forces from the magnetic field. However, no mechanism provides super-Keplerian support of the flow in this region. The ions still lag behind the neutrals, providing an azimuthal drag which decelerates the neutrals (WK93). For the flow to remain sub-Keplerian within the disc, the configuration must satisfy
\begin{equation}
\frac{{\rm d} B_{\rm r}}{{\rm d} B_{\rm \phi}} = - \frac{J_{\rm \phi}}{J_{\rm r}} \approx - \frac{\left(\etaHnorm + 2 \right) B_{\rm z}^2 + \etaAnorm B_{\rm r} B_{\rm \phi}}{\etaOnorm B_{\rm z}^2 + \etaAnorm \left( B_{\rm r}^2 + B_{\rm z}^2\right)} < 0 \label{eqn:const1}
\end{equation}
below the disc surface (see Section 4.1 of KSW10).

\item[(ii)] \textit{Wind launching}: A wind is driven from the disc surface (i.e. a wind launching criterion is satisfied). As shown by BP82 in the ideal-MHD limit, a minimum inclination angle is required between the surface magnetic field (denoted by the subscript $b$) and the rotation axis to launch a wind. In Section 4.2 of KSW10, this constraint was generalized for non-ideal MHD to the following:
\begin{eqnarray}
\left[ 3 + \frac{3}{2} \etaHnorm - \etaAnorm \tilde{\eta}_{\rm P} \right] B_{\rm rb}^2 &>& \left[1 + \frac{5}{2} \etaHnorm + \tilde{\eta}_{\rm T}^2 \right] B_{\rm zb}^2 \nonumber \\
&& {}+ \etaAnorm \tilde{\eta}_{\rm P} B_{\rm \phi b}^2 + \frac{3}{2} \etaOnorm B_{\rm rb} B_{\rm \phi b}, \label{eqn:const2}
\end{eqnarray}
below the disc surface, where $\tilde{\eta}_{\rm T}^2 \equiv \etaOnorm^2 + \etaHnorm^2 + \etaAnorm^2$ and \mbox{$\tilde{\eta}_{\rm P} \equiv \etaOnorm + \etaAnorm$}. This is used as a necessary condition for wind launching in our models. If this constraint is not satisfied, then there is either insufficient field-matter coupling to bend the magnetic field lines past the critical angle required for wind-launching, or the magnetic field is too strong to be bent. In the ideal-MHD limit, equation (\ref{eqn:const2}) reduces to the BP82 wind-launching criterion $B_{\rm rb}/B_{\rm zb} > 1/\sqrt{3}$.

\item[(iii)] \textit{Mass loading}: Only the upper layers of the disc participate in the outflow. According to both theoretical and observational arguments \citep[e.g.][]{Konigl:2000tf}, only a small fraction of the disc material should participate in the outflow. If the wind torque is too strong, the disc wind is inherently unstable \citep{Cao:2002cc}. The mass loading condition is implemented in KSW10 by requiring that the base of the wind $\zb$ (which we identify as the height above which $\vphi > \vk$) be located above the magnetically reduced density scale height $\zh$, defined as the height at which the density drops to $\rhoO / \sqrt{e}$,
\begin{equation}
\zb > \zh. \label{eqn:const3a}
\end{equation}
WK93 showed that if equation (\ref{eqn:const3a}) is not satisfied, the gradient of $B_{\rm \phi}$ changes sign within the disc. They explained this behaviour by noting that, as the midplane inflow Mach number $\epsilon$ decreases, the normalized height of the sonic point $\tilde{z}_s$ decreases, and the scale height increases. Eventually, $\rhosnorm = \rho_{\rm s}/\rhoO$ becomes so large that the upwards mass flux transports more angular momentum than that brought in by the accretion flow. As a result, the gradient of $B_{\rm \phi}$ changes sign as the magnetic field begins transferring angular momentum back into the flow before the top of the disc is reached. Such a configuration is unphysical, and likely unstable. Since the gradient of $B_{\rm \phi}$ changes sign to a small degree below the sonic point in all of our 1.5D solutions, we have devised a new constraint based on the magnitude of this change.

In practice we expect that magnetic energy dominates up to the Alfv\'{e}n surface, occuring at heights of order the footpoint radius, for $r_{\rm A}/r_{\rm 0} \sim 3$ and typical inclinations of the magnetic field. This is also the scale at which collimation takes place. Hence, we define a new constraint for the mass loading by requiring that the extrapolated value of $B_{\rm \phi}$ at $z = r$ be less than zero, i.e.
\begin{equation}
B_{\rm \phi}(z = r) < 0. \label{eqn:const3}
\end{equation}
Due to the high sensitivity of the radial splitting and ejection-accretion ratio of the wind-launching region to this constraint, we treat all consequential results with caution, noting that our current models are not able to follow $B_{\rm \phi}$ beyond $\zs$. Thus, the mass loading constraint is only an approximate constraint that results from extrapolating the solution to $z > \zs$ and therefore, the radial location where it is violated is approximate.

\item[(iv)] \textit{Energy conservation}: The rate of heating by Joule dissipation at the midplane is bounded by the rate of gravitational potential energy released at that location \citep{Konigl:1997tu},
\begin{equation}
(\mathbf{J} \cdot \mathbf{E}')_{\rm 0} < \frac{\epsilon \vk}{2 \aO^2}\frac{B_{\rm 0}^2}{4 \pi \hthermal}. \label{eqn:const4}
\end{equation}
\end{description}
Diffusion-regime specific simplifications of each of these constraints may be found in WK93 (Ambipolar regime) and KSW10 (Hall and Ohm regimes). Here we use the generalized form which applies to all three regimes.

%----------------------------------------------
% 1+1.5D WIND-DRIVING DISC SOLUTIONS
%----------------------------------------------
\section{1+1.5D disc wind models} \label{sec:2d}

Having established the basis for the 1.5D solutions, we now discuss the 1+1.5D framework in detail. We begin by describing the parameters which define the structure of the disc, and then outline the method used to find each 1.5D solution that matches this structure, in order to build the 1+1.5D wind-launching model. Finally, we provide an overview of the requirements for such a model, in the interest of self-consistency and physical viability.

\subsection{1+1.5D parameters} \label{sec:2d:param}
To build a 1+1.5D model, a framework must be constructed which connects the 1.5D solutions together in a physically consistent way. In our model, this framework is constructed from four disc parameters. The first three are defined as analytic functions of radius, while the fourth depends primarily on a single value at the innermost radius, and is calculated for all other radii using mass conservation. We now describe each parameter in detail. 

%----- Magnetic field strength -----
\paragraph*{Magnetic field strength:} \label{sec:mag_struct}
The ratio of the Alfv\'{e}n speed to the isothermal sound speed at the disc midplane \mbox{$\aO$ ($\equiv \vAO/\cs$)} is a measure of the magnetic field strength. We assume that $\aO$ is constant for the entire wind-launching region of the disc, and consider values of $\aO \sim 1$ in this study. This parameter is bounded for wind solutions, since weaker magnetic fields ($\aO \ll 1$) cause the MRI to dominate and drive redistribution of angular momentum. On the other hand, stronger magnetic fields ($\aO \gtrsim 1$) inhibit wind launching due to their stiffness (KSW10). 

%----- Temperature -----
\paragraph*{Temperature:} \label{sec:temp}
We assume that the disc is vertically isothermal and prescribe the radial temperature profile via the minimum mass solar nebula (MMSN) prescription \citep{Hayashi:1981ws, Hayashi:1985vv},
\begin{equation}
T(r) = T_{\rm 0} \left(\frac{r}{1 \mbox{ au}} \right)^{-q}, \label{eqn:temp}
\end{equation}
for which $T_{\rm 0} = 280$ K and $q = 0.5$. Recent observations confirm $q = 0.5$ as a reasonable value for circumstellar discs \citep{Andrews:2005to, Andrews:2007hv}. The disc temperature is a complex function of key parameters, such as composition, abundance and properties of dust grains, density, ionization state, disc activity, chemistry, as well as the penetration of external radiation fields (X-rays, cosmic rays, and stellar irradiation).

%----- Surface density -----
\paragraph*{Surface density:} \label{sec:surfdens} For the surface density, we adopt a radial power-law dependence of the form
\begin{equation}
\Sigma (r) = \SigmaO \left(\frac{r}{1\mbox{ au}}\right)^{-p}, \label{eqn:surfdens2}
\end{equation}
similar to the MMSN formulation, where the surface density at any radius is defined by 
\begin{equation}
\Sigma = 2 \int^{\zs}_{\rm 0} \rho {\rm d}z.\label{eqn:surfdens}
\end{equation}
$\Sigma(r)$ directly influences the amount of ionizing radiation that reaches the disc midplane. This in turn controls the ionization balance, and resulting conductivity structure, which governs the dynamics and evolution of the disc. The radial surface density structure of a purely wind-driving disc is expected to be flatter and thinner than a MMSN disc \citep{Combet:2008cm}, however according to observations, both $\SigmaO$ and $p$ have large ranges \citep[see][]{Andrews:2007hv, Persson:2016dd}. This is taken into account in the present models by choosing a range of values for $\SigmaO$ that are lower than that of the MMSN, as well as flatter radial profiles (lower values of $p$).

%----- Mass accretion rate -----
\paragraph*{Mass accretion rate:} \label{sec:mass_accr} 
We define the `local' mass accretion rate as
\begin{equation}
\mdota (r) = -2 \pi r \int^{\zb}_{-\zb} \rho \vr {\rm d}z. \label{eqn:mdot_a}
\end{equation}
This measures the mass of material falling inward through a disc annulus centred at radius $r$ per unit time. Using equation (\ref{eqn:mdot_a}), we define the mass accretion rate at the inner radius of the wind-launching region ($\rin$) as
\begin{equation}
\mdotin = \mdota(\rin). \label{eqn:mdotin}
\end{equation}
Following this, we calculate $\mdota$ at larger radii $r$ by adding the wind mass flux between $\rin$ and $r$ to $\mdotin$ (see Section \ref{sec:2d:build}). The accretion rate and wind mass flux are derived by vertical integration of the steady-state mass conservation law in cylindrical coordinates
\begin{equation}
\frac{1}{r}\frac{\partial}{\partial r}\left(r \rho \vr \right) + \frac{\partial}{\partial z} \left( \rho v_z \right) = 0. \label{eqn:continuity}
\end{equation}
This integration is performed between $-\zb$ and $\zb$, where $\zb$ is the vertical height of the disc surface (the height above which $\vphi > \vk$), since material begins to move radially outward via the magnetocentrifugal mechanism above $\zb$. Thus we obtain 
\begin{equation}
\frac{{\rm d}}{{\rm d}r} \int^{\zb}_{-\zb}2 \pi r \rho \vr {\rm d}z + 4 \pi r \rhob v_{\rm zb} = 0. \label{eqn:ddzmassconserv}
\end{equation}
Combining equation (\ref{eqn:mdot_a}) with the wind mass loss rate
\begin{equation}
\mdotw (r) = 4 \pi \int^{\rout}_{r} r' \rhob v_{\rm zb} {\rm d}r', \label{eqn:mdot_w}
\end{equation}
where $\rout$ is the outer radius of the wind-launching region, equation (\ref{eqn:ddzmassconserv}) can be rewritten as
\begin{equation}
\frac{{\rm d}}{{\rm d}r}\mdota (r) = 4 \pi r \rhob v_{\rm zb} = -\frac{{\rm d}}{{\rm d}r}\mdotw (r). \label{eqn:ddrlocmass}
\end{equation}
This implies 
\begin{equation}
\dot{M} = \mdota (r) + \mdotw(r) = \mbox{const} \label{eqn:totmassflux}
\end{equation}
\citep[e.g.][]{Kuncic:2004tf}, where $\dot{M}$ is the total mass flux at large radii. Hence the radial profiles of mass accretion rate and wind mass loss rate are inextricably linked.

In addition to the quantities listed above, we define the cumulative wind mass loss rate over the entire wind-launching region as
\begin{equation}
\mdotout = \mdotw(\rin). \label{eqn:mdotout}
\end{equation}
The ratio $\mdotout/\mdotin$ is the ejection/accretion ratio, which is a key observational parameter for constraining the acceleration mechanism in protostellar disc winds and jets. For disc winds, the one-sided ejection/accretion ratio ($\mdotout/2\mdotin$) is predicted to be 0.1 \citep{Pelletier:1992fk}, with recent observations confirming this \citep[e.g.][]{Cabrit:2007hz, Ellerbroek:2013bf, Watson:2015uo}. In Section \ref{sec:change_var}, we determine the dependence of $\mdotout/\mdotin$ on the accretion rate $\mdotin$ and the radial profiles of the magnetic field strength via $\aO(r)$, and the surface density $\Sigma(r)$.

%---------- Constructing the disc model ----------
\subsection{Constructing the disc model} \label{sec:2d:build}

\begin{table*}
\caption{A listing of both the parameters used to describe the 1+1.5D solutions and those that describe the 1.5D radially localized wind-driving disc solutions.}
\begin{center}
\begin{tabular}{clcl}
\hline
\multicolumn{2}{|c|}{1+1.5D parameters} & \multicolumn{2}{|c|}{1.5D parameters} \\
\hline
$\aO(r)$ & Radial profile of the ratio $\vA/\cs$ at the disc midplane & $\aO$ & Ratio $\vA/\cs$ at the disc midplane\\
$T(r)$ & Radial isothermal disc temperature profile & $\cs/\vk$ & Geometric disc thickness ratio \\
$\Sigma(r)$ & Radial surface density profile & $\left[\etaH/\etaO\right]_{\rm 0}$ & Midplane Hall-to-Ohm diffusivity ratio \\
$\mdotin$ & Mass accretion rate at the inner radius $\rin$ & $\left[\etaA/\etaO\right]_{\rm 0}$ & Midplane ambipolar-to-Ohm diffusivity ratio\\
& & $\ElsasserO$ & Midplane field-neutral coupling parameter \\
&& $\epsilon$ & Normalized inward radial speed at the disc midplane \\
\hline
\end{tabular}
\end{center}
\label{tab:variables}
\end{table*}

Now that we have set the 1+1.5D framework, we can begin building our model from 1.5D solutions. In order to convey the method clearly, we first describe how we arrive at a 1.5D solution for any given combination of values for the set of four 1+1.5D parameters $\aO$, $T$, $\Sigma$, and $\mdota$. We then outline our procedure for finding both $\rin$ and its corresponding local solution given these values, and conclude with our approach for extending the model outward from $\rin$ and how we determine the outer edge of the wind-launching region $\rout$.

For any given combination of values for $\aO$, $T$, $\Sigma$, and $\mdota$, defined at a particular radius $r$, and for a stellar mass $\mstar$, there may exist a unique local solution which satisfies these values. Each 1.5D solution is characterized by six parameters (see Table \ref{tab:variables} and Section \ref{sec:1.5D:parameters} for more detail), and each of these local parameters must be derived from the 1+1.5D parameters (including $r$ and $\mstar$) in order to calculate the matching 1.5D solution. While $\aO$, $T$, $r$ and $\mstar$ are used to directly calculate the local parameters, $\Sigma$ and $\mdota$ may only be determined once the local solution is known. Therefore, we begin by estimating the values of $\rhoO$ and $\vrO$, the midplane density and radial velocity respectively, and evaluate $\Sigma$ and $\mdota$ from the resulting solution. We then adjust the values of $\rhoO$ and $\vrO$ accordingly, and iterate on them until the resulting $\Sigma$ and $\mdota$ match their assigned values to within $10^{-6}$. Using this method, we can now find the inner radius of the wind-launching region and its corresponding solution.

To find $\rin$ and its solution, we begin by specifying the radial profiles of $\aO(r)$, $T(r)$, $\Sigma(r)$ and the accretion rate at the inner edge of the wind region, $\mdota = \mdotin$ (we vary these profiles in Section \ref{sec:change_var} in order to measure the dependency of the wind-launching region on them). We then search for the innermost radius which satisfies these requirements while being physically viable (see Section \ref{sec:1.5D:const} for a detailed description of the constraints which determine whether a solution is physically viable). This involves first finding a valid solution at any radius that satisfies the 1.5D constraints, and then stepping inwards in $r$ until solutions become invalid. We then take the smallest radius which gives a valid solution and designate it $\rin$, and its solution becomes the basis for constructing the rest of the 1+1.5D model.

Once the inner radius and its solution are known, we calculate solutions at logarithmically increasing intervals of radius until they are no longer valid. These discrete intervals are defined by
\begin{equation}
r_{\rm i+1} = 10^{1/k} r_{\rm i}, \label{eqn:k}
\end{equation}
or
\begin{equation}
\Delta r_{\rm i} = r_{\rm i+1} - r_{\rm i} = (10^{1/k} - 1) r_{\rm i}, 
\end{equation}
where $k$ is the number of 1.5D solutions per decade of radius. We choose $k = 1000$ for all models described in this paper based on a numerical convergence study, which is included in Appendix \ref{sec:res}. 

The 1+1.5D profiles $\aO(r)$, $T(r)$ and $\Sigma(r)$ are already defined for all $r$, so that the only parameter remaining to calculate at each new radial step $r_{\rm i} + \Delta r$ is the local mass accretion rate. This is determined by adding the wind flux in the interval $[r_{\rm i}, r_{\rm i} + \Delta r]$ to the local accretion rate at $r_{\rm i}$,
\begin{eqnarray}
\mdota (r_{\rm i} + \Delta r) &=& \mdota (r_{\rm i}) + \mdotw (r_{\rm i}) - \mdotw(r_{\rm i} + \Delta r) \nonumber \\
&=& \mdota (r_{\rm i}) + 4 \pi \int^{r_{\rm i} + \Delta r}_{r_{\rm i}} r \rhob (r) v_{\rm zb} (r) {\rm d}r \nonumber \\
&\simeq& \mdota (r_{\rm i}) + 4 \pi r_{\rm i} \rhob (r_{\rm i}) v_{\rm zb} (r_{\rm i}) \Delta r
\end{eqnarray}
(see equation \ref{eqn:ddrlocmass}). This process is repeated until we reach a radius where the solution is no longer valid, thus defining the outer radius of the wind-launching region, $\rout$.

%---------- Constraints on the disc model ----------
\subsection{Constraints on the disc model} \label{sec:2d:const}

In order for our 1+1.5D models to be self-consistent, they must not violate the equations and assumptions on which the individual 1.5D solutions are based. These solutions are described in detail by WK93 and KSW10, and we summarized their main characteristics in Section \ref{sec:1.5D}. We now discuss how we preserve the assumptions of each solution, and other continuity properties within our 1+1.5D model.

The primary assumption employed by KSW10 in their 1.5D models is that of geometrical thinness, which permits neglecting the radial derivative terms ($|\partial/\partial r| \sim 1/r$) in the equations of MHD in comparison with vertical derivative terms ($|\partial/\partial z| \sim 1/\zh$), where $\zh$ ($\ll r$) is the disc density scaleheight. In Sections 3.2--3.10 of KSW10, this assumption simplifies the axisymmetric equations of non-ideal MHD into a set of six ODEs in $z$. For each of the 1+1.5D models, we ensure that the disc remains geometrically thin at all radii, thereby fulfilling this requirement.

The 1+1.5D models must also satisfy $\nabla \cdot \mathbf{B} = 0$ to prevent an unrealistic magnetic field configuration. In the axisymmetric limit, this constraint is:
\begin{equation}
\frac{1}{r}\frac{\partial}{\partial r}\left(r B_{\rm r} \right) + \frac{\partial}{\partial z} \left( B_{\rm z} \right) = 0, \label{eqn:divb}
\end{equation}
and by neglecting the radial derivative, equation (\ref{eqn:divb}) implies that $B_{\rm z}$ is constant with height. This result is used in the 1D solution.

In order to adequately satisfy the divergence constraint in the 1+1.5D models, it is sufficient to show that the scale height of $B_{\rm z}$ implied by the radial component of the magnetic flux density, $B_{\rm r}(r,z)$ is much larger than the disk scale height. To check this, we take the profile of $B_{\rm r}(r,z)$, calculated by interpolating $B_{\rm r}$ over the extend of the completed 1+1D model in ($r$, $z$) space, calculate $r^{-1} \partial (r B_{\rm r})/\partial r$ and then use equation (\ref{eqn:divb}) to estimate a local scale length, $L$, for $B_{\rm z}$, that is,
\begin{equation}
L = B_{\rm z} \left( \frac {\partial B_{\rm z}}{\partial z} \right)^{-1}. \label{eqn:l}
\end{equation}
If $L \gg \zh$ then the $\nabla \cdot \mathbf{B} = 0$ condition is adequately met. We have checked that this condition is satisfied at all points in all of our 1+1.5D models, and include results for $L$ in Appendix \ref{sec:divb} for the strong-wind model described in Section \ref{sec:comp:strong}.

%----------------------------------------------
% FIDUCIAL MODELS
%---------------------------------------------- 
\section{A comparison of weak and strong wind-driving discs} \label{sec:comp}

%Summary and technical details
We now examine in detail the internal structure of the wind-launching region in protostellar discs for two distinct cases. We prescribe both discs with a surface density profile $\Sigma(r) = 630 (r/$au$)^{-1.0}$ g cm$^{-2}$, and a magnetic field strength $B_{\rm z}$ corresponding to $\aO = 1.0$. We purposely choose the surface density constant $\SigmaO$ and power-law index $p$ (see equation \ref{eqn:surfdens2}) to be lower than the MMSN values of $\SigmaO = 1700$ g cm$^{-2}$ and $p = 1.5$ respectively, because of the expectation that the radial surface density structure of a purely wind-driving disc is thinner and flatter than a MMSN disc \citep{Combet:2008cm}. However, we investigate the effect of changing $\SigmaO$ and $p$ in Section \ref{sec:change_var}. 

We prescribe the first model with an accretion rate of $\mdotin = 1.0 \times 10^{-5} \msun$ yr$^{-1}$, and the second with \mbox{$\mdotin = 1.6 \times 10^{-5} \msun$ yr$^{-1}$}, then calculate the extent of the 1+1.5D model via the procedure described in Section \ref{sec:2d}. We find that the first model exhibits a weak wind \mbox{($\mdotout/\mdotin = 1.3 \times 10^{-2}$)} relative to the second \mbox{($\mdotout/\mdotin = 3.5 \times 10^{-2}$)} and hence we refer to them as the weak and strong wind models for the remainder of this paper. 

%---------- Weak-wind model ----------
\subsection{Weak-wind model} \label{sec:comp:weak}

\begin{figure*}
	\centering
	\includegraphics[width=177mm]{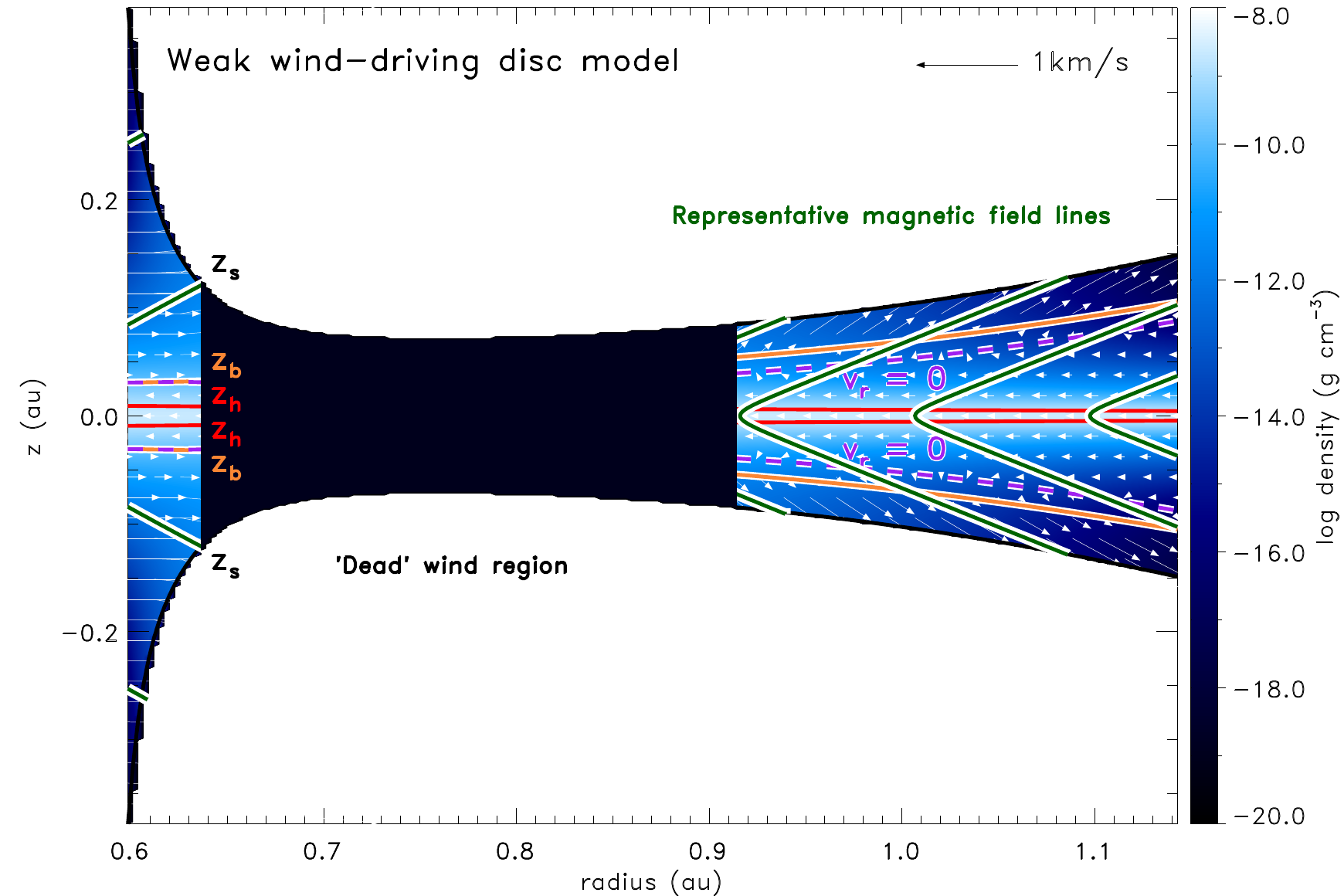}
	\caption{A poloidal slice of the weak-wind (lower ejection/accretion ratio) disc model. The density contour plot is overlaid with velocity vectors in white and green lines to show the bending of magnetic field lines. The black area defines the region where wind solutions are unphysical according to the mass-loading constraint; the `inactive' region (see Section \ref{sec:1.5D:const}). The red lines indicate the magnetically reduced disc scale height $\zh$ and the orange lines show the location of the disc surface/base of the wind $\zb$, defined as the height at which the azimuthal velocity of the gas transitions from sub-Keplerian ($z < \zb$) to super-Keplerian ($z > \zb$). The black lines indicate the sonic surface $\zs$. The purple dashed lines indicate the extents of the accretion region within the disc, where $v_{\rm r} < 0$.}
	\label{fig:low_disc_rho}
\end{figure*}

%Extents of the disc.
The density, velocity and magnetic field structure of the weak-wind model is displayed in Fig. \ref{fig:low_disc_rho}. The magnetic field lines in Figs. \ref{fig:low_disc_rho} and \ref{fig:high_disc_rho} are calculated from their vertical and radial components, where all three components of the magnetic field ($B_{\rm r}$, $B_{\rm \phi}$, $B_{\rm z}$) are calculated in each 1.5D vertical model. To make the 2D images in Figs. \ref{fig:low_disc_rho} and \ref{fig:high_disc_rho}, we simply interpolate the $B_{\rm r}$ and $B_{\rm z}$ components in ($r$, $z$) space. We use a similar procedure for the velocity field, and perform a basic interpolation for the density.

The wind-launching region of the disc is radially localized, and extends from $\sim0.60$ to 1.14 au. However, between $\sim0.63$ and 0.92 au there exists a region, which the mass loading constraint defines as ineffective to wind launching, which we denote the `inactive' region (see Section \ref{sec:1.5D:const}). Essentially, the extrapolated $B_{\rm \phi}$ in this region changes sign below $z = r$, signalling an unphysical transfer of angular momentum from the field back to the matter. Hence a stable CDW cannot operate within this radial range, and we do not include the contribution of the solutions in this region to the total wind mass loss rate $\mdotout$. Despite there being no capacity to launch a stable wind in the inactive region, this does not rule out unstable launching configurations such as episodic outbursts from existing here, however in order to verify this and determine the dominant mode of angular momentum transport in this region, a time-dependent treatment of the disc is necessary, which is beyond the scope of this paper.

It is worth noting that despite the appearance of a sharp cutoff for the inactive region in Fig. \ref{fig:low_disc_rho}, the assumptions used in formulating the mass-loading constraint are approximate. Hence the radial extent of this region is to be treated as a first approximation, and likewise the values for the total bipolar wind mass-loss rate, as they are directly connected to the radial extent of the wind-launching region.

The accretion region within the disc is the zone where the flow proceeds inwards ($v_{\rm r} < 0$). The boundaries of this region are marked by the purple dashed lines in Fig. \ref{fig:low_disc_rho}. At the inner wind radius ($r = 0.6$ au), the accretion region is located between $\pm 0.031$ au, which is, equivalently, $\pm 3.3$ $\zh$ and $\pm 1.8$ $\hthermal$. Similarly, at the outer wind radius ($r = 1.14$ au), the accretion region is located between $\pm 0.088$ au, equivalently $\pm 19$ $\zh$ and $\pm 2.2$ $\hthermal$.

Note that the disc in Fig. \ref{fig:low_disc_rho} exhibits the three distinct layers described by \citet{Konigl:2011tk}. The \textit{quasi-hydrostatic region} straddles the disc midplane ($|\rm{z}| < \zh$), and is matter dominated; the majority of mass accretion and magnetic field bending and shearing take place in this region (as apparent by the white velocity vectors and overlaid magnetic field lines in green). Above the quasi-hydrostatic layer lies the \textit{transition zone} ($\zh < |z| < \zb$). In this layer the magnetic field lines become locally straight as the density decreases and the flow becomes magnetically dominated. The inward radial flux of matter gradually decreases, and the flow transitions to a CDW (as shown by the white velocity vectors). At the top of the transition zone the flow changes from sub-Keplerian ($\vphi < \vk$) to super-Keplerian ($\vphi > \vk$). This point represents the base of the wind ($\zb$), above which lies the \textit{outflow region} ($\zb < |z| < \zs$). Here the flow continues to accelerate until it reaches the sonic surface ($\zs$) which defines the extent of the model.

The mass accretion rate through the inner boundary for the weak-wind model is $\mdotin = 1.0 \times 10^{-5} \msun$ yr$^{-1}$, with a total bipolar-wind mass loss rate $\mdotout = 1.3 \times 10^{-7} \msun$ yr$^{-1}$. This equates to an ejection/accretion ratio $\mdotout/\mdotin = 1.3 \times 10^{-2}$, which is approximately an order of magnitude lower than the observationally inferred average of 0.1--0.2, but nonetheless it is exhibited by some protostellar systems \citep[see][Fig. 1]{Cabrit:2007hz}. These and other properties of the disc at the inner and outer radii of the wind-driving region, such as the Ohm, Hall and ambipolar diffusivities and scale heights, are listed in Table \ref{tab:fiducial}.

%---------- Strong-wind model ----------
\subsection{Strong-wind model} \label{sec:comp:strong}

\begin{figure*}
	\centering
	\includegraphics[width=177mm]{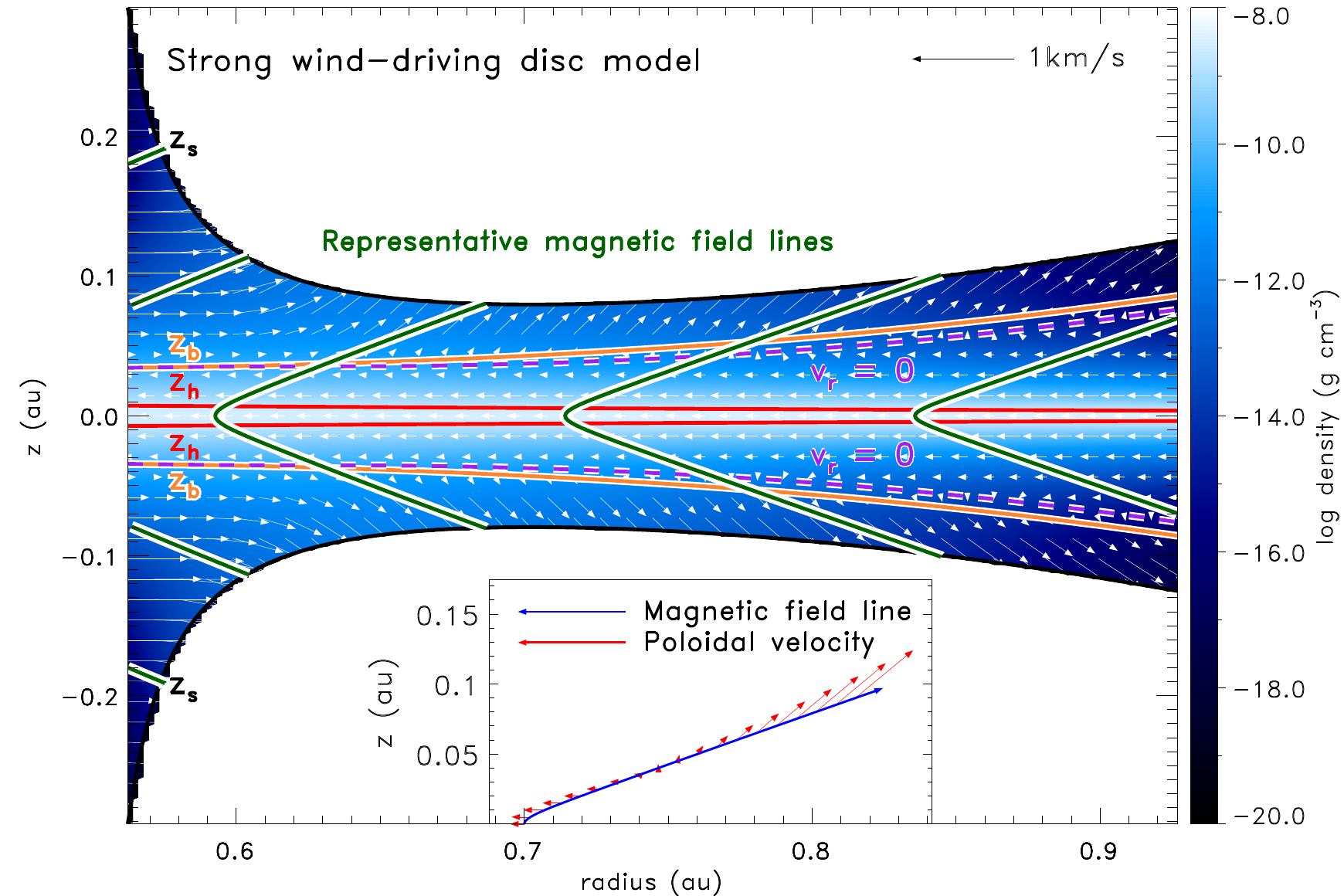}
	\caption{A poloidal slice of the strong-wind disc model. See Fig. \ref{fig:low_disc_rho} for a description of the features of this plot. In addition, this plot includes an inset, comparing a sample magnetic field line anchored at 0.7 au with the velocity field along its length. We discuss the features of this comparison in Section \ref{sec:comp:strong}.}
	\label{fig:high_disc_rho}
\end{figure*}

The strong-wind model (Fig. \ref{fig:high_disc_rho}) has a similar radial profile to the weak-wind model, with an elevated disc surface ($\zb$) and sonic surface ($\zs$) at small radii, however this model does not possess an internal magnetically inactive region. The wind-launching region of the disc extends from $\sim 0.56$ to $0.93$ au, and is located closer to the star than that of the weak-wind model discussed above. Similar to the weak-wind model, the wind-launching region is localized, with maximum wind output at $r = 0.65$ au, and rapid attenuation of wind output on either side (see Section \ref{sec:comp:rates}).

At the inner wind radius of the strong-wind model ($r = 0.56$ au), the accretion region is located between $\pm 0.034$ au, or equivalently, $\pm 4.7$ $\zh$ and $\pm 2.1$ $\hthermal$. At the outer wind radius, this region expands to $\pm 0.076$ au, or $\pm 21$ $\zh$, and $\pm 2.5$ $\hthermal$.

The mass accretion rate through the inner boundary of this model is $\mdotin = 1.6 \times 10^{-5} \msun$ yr$^{-1}$, with a total bipolar-wind mass loss rate of $\mdotout = 5.6 \times 10^{-7} \msun$ yr$^{-1}$. This results in an ejection/accretion ratio $\mdotout/\mdotin = 3.5 \times 10^{-2}$, which is also lower than the mean observed range of $\simeq 0.1$--0.2 for protostellar discs, but some individual observed discs show such low values \citep{Cabrit:2007hz, Ellerbroek:2013bf, Watson:2015uo}. This value is a factor of 2.7 larger than that of the weak-wind model. The relation between $\mdotin$ and $\mdotout$ as a function of 1+1.5D disc properties is explored in further detail in Section \ref{sec:change_var}.

In the inset of Fig. \ref{fig:high_disc_rho}, we compare the angle between the magnetic and velocity field vectors along a single magnetic field line. As expected, the magnetic field begins at right angles to the flow, and is bent outward, consistent with the flow of neutrals towards the centre of the system, and the support of matter against gravity by magnetic tension, given that the flow is sub-Keplerian. Moving up to the transition zone, the field line becomes locally straight as the flow transitions to being magnetically dominated. This causes the velocity field to gradually align with the magnetic field as it moves away from the disc midplane. It is expected that the two fields will eventually align perfectly when moving into the ideal MHD regime. This is not seen in our models, because we do not follow the wind solutions into the regions of near-ideal MHD, and in particular, our current model does not account for the physical dependence of the Elsasser number on $z$ (which will be addressed in a subsequent paper).

%---------- Radial dependence of mass accretion and outflow rates ----------
\subsection{Radial dependence of mass accretion and outflow rates} \label{sec:comp:rates}

\begin{figure}
	\centering
	\includegraphics[width=84mm]{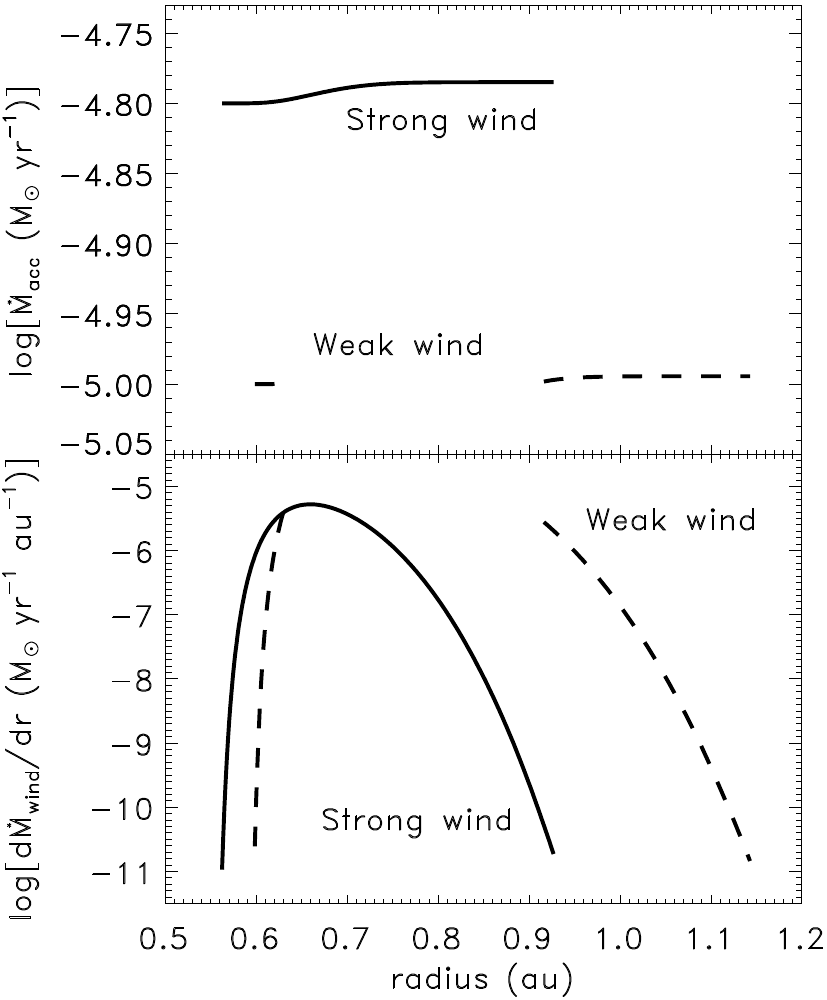}
	\caption{Local mass accretion rate $\mdota$ (top panel) and wind mass loss rate per unit radius ${\rm d}\mdotw/{\rm d}r$ (bottom panel) as a function of radius for the weak-wind (dashed line) and strong-wind (solid line) models displayed in Figs. \ref{fig:low_disc_rho} and \ref{fig:high_disc_rho} respectively.}
	\label{fig:low_high_mdota}
\end{figure}

The radial dependence of the local mass accretion rate $\mdota$ and wind mass loss rate per unit radius ${\rm d}\mdotw/{\rm d}r$ for the weak-wind and strong-wind models are shown as the dashed and solid lines in Fig. \ref{fig:low_high_mdota}, respectively. For the weak-wind model, the local wind outflow rate increases towards intermediate radii, but above \mbox{${\rm d}\mdotw/{\rm d}r \approx 3\times 10^{-6} \msun$ yr$^{-1}$ au$^{-1}$} the wind solutions become unphysical and the disc becomes ineffective to wind-launching. For the strong-wind model, the region of highest wind mass loss is between $0.6 < r < 0.8$ au, peaking at $r = 0.65$ au. 

To understand why the outflow is radially localized, we break down the explanation into two parts:
Firstly, we explain the decrease in the wind outflow rate towards smaller radii in the inner section of the wind-launching region. This is primarily a result of weakened coupling between the field and the matter in the inner regions of the disc, with the midplane Elsasser number approaching $\ElsasserO = 1$ towards the inner limit of the wind region, from a value of $\ElsasserO = 12$ at the outer limit (see Table \ref{tab:fiducial}). As shown by KSW10, disc properties behave quite differently for $\ElsasserO \lesssim 1$. Referring to the description of a CDW mechanism by \citet{Konigl:2011tk}, decreased coupling between the field and the neutrals leads to the following effects: the azimuthal velocity $\vphi$ increases as the magnetic torque diminishes and as such, the inward flow of neutrals decreases. This reduces the radial drag on the magnetic field lines, contributing to a decrease in $B_{\rm r}/B_{\rm z}$, which reduces the magnetic compression of the disc and results in lower density stratification. The angle between the surface magnetic field and the rotation axis is critical to launching a wind (BP82), hence, as $B_{\rm r}/B_{\rm z}$ decreases the local wind mass loss rate ${\rm d}\mdotw (r)/{\rm d}r \rightarrow 0$. In effect, this drop in wind mass loss rate supports the validity of the simplified wind launching criterion for non-ideal MHD (equation \ref{eqn:const2}), for none of the models in this paper actually violate equation (\ref{eqn:const2}), as the solutions approach the inward radius where this criterion would be violated, the wind drops to negligible levels.

The increase in the height of the sonic surface with decreasing $r$ for $r < 0.7$ au can also be explained via the field-matter coupling. The decrease in $B_{\rm r}/B_{\rm z}$ combined with lower coupling reduces the extraction of angular momentum via the wind, and therefore the vertical distance required for $\vz$ to reach the sound speed ($\zs$) grows. Hence the height of the sonic surface above the disc increases as the field-matter coupling weakens.

Secondly, we explain the decrease in the wind outflow rate towards larger radii in the outer region of the wind. In this region, the field-matter coupling continues to increase with radius (see Fig. \ref{fig:elsasser}). This reduces the azimuthal velocity $\vphi$ as the magnetic torque increases, increasing the inward flow of neutrals and leading to greater field bending in both the radial and azimuthal directions (i.e. the ratios $B_{\rm r}/B_{\rm z}$ and $B_{\rm \phi}/B_{\rm z}$). However, even though these are favourable conditions for wind launching, another effect becomes important here which counters the launching of a wind. Due to the increased field bending at larger radii, magnetic compression of the disc material results in a lower density at the disc surface, producing a lower density wind with a lower mass loss rate.
This is amplified by the rising of the disc surface $\zb$ with radius, leading to an even lower density wind. Since the azimuthal velocity at the disc midplane ($v_{\rm \phi 0}$) decreases with radius as a result of increased field-matter coupling, the vertical distance required for $\vphi$ to reach the Keplerian speed ($\vk$) grows. Hence the disc surface, defined as the height at which $\vphi = \vk$, rises with radius, and as a result the wind density decreases, and the outflow rate becomes negligible. Thus the wind-launching region is radially localized as a consequence of two independent mechanisms, one operating at inner radii, and the other at outer radii.

%---------- Comparison with other models ----------
\subsection{Comparison with other models} \label{sec:comp:diff}

Comparing our work to the models of \citet{Bai:2013ue}, we find some significant differences. Firstly, the values of $\zb/\hthermal$ are twice as large as those found in our study over the entire radial range. In \citet{Bai:2013ue} the models are centred at 1 au, and give values of $\zb/\hthermal \sim 4.6$, whereas those found in our two fiducial models have values between 1.8 and 2.7 for the weak wind, and 2.1--2.8 for the strong wind. This difference is caused by the vastly different magnetic field strengths in both simulations. In our models, the midplane magnetic field is $10^5$ times stronger, leading to much larger compressive forces on the disc.

Secondly, the field morphologies of both models are completely different. The models of \citet{Bai:2013ue} contain a laminar region near the disc midplane, associated with relatively little field-line bending due to the strong diffusion and weak field strength (see their Figure 6). On the other hand, the field lines in our models begin to bend immediately above the disc midplane as expected (see inset to Fig. \ref{fig:high_disc_rho}), given the relatively large field strength.

\begin{figure}
	\centering
	\includegraphics[width=84mm]{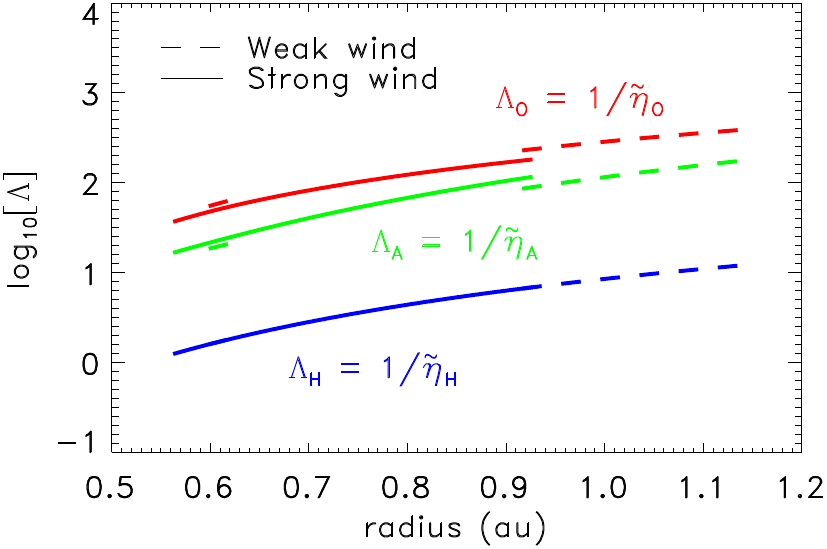}
	\caption{Radial profile of the Ohmic ($\Lambda_{\rm O}$), Hall ($\Lambda_{\rm H}$), and ambipolar ($\Lambda_{\rm A}$) Elsasser numbers, for both the weak and strong wind models (given the simplifications of the model, Elsasser numbers are constant with $z$).}
	\label{fig:elsasser}
\end{figure}

Thirdly, \citet{Bai:2013ue} find that both the Ohmic and ambipolar Elsasser numbers must exceed unity in order to facilitate wind launching, which means that the total Elsasser number must remain above unity. Due to the simplifications of our model, the Elsasser numbers remain constant with height, however we can still check the dependence of this parameter on radius. Fig. \ref{fig:elsasser} displays the three Elsasser numbers for both fiducial models as a function of radius. We find agreement with \citet{Bai:2013ue}, in that wind launching occurs where both $\Lambda_{\rm O}$ and $\Lambda_{\rm A}$ exceed unity. Hence we find it to be a necessary but not sufficient condition for wind launching.

\begin{table*}
\caption{Properties of the weak and strong-wind disc models presented in Section \ref{sec:comp}. These properties are given at the inner ($\rin$) and outer ($\rout$) radial extents of the wind-launching region.}
\begin{center}
\begin{tabular}{clllllll}
\hline
Disc property & & \multicolumn{4}{|c|}{Weak-wind model} & \multicolumn{2}{|c|}{Strong-wind model} \\
&& \multicolumn{2}{|c|}{Inner region} & \multicolumn{2}{|c|}{Outer region} &&\\
& & $\rin$ & $\rout$ & $\rin$ & $\rout$ & $\rin$ & $\rout$ \\
 \hline
$r$ & [au] & 0.60 & 0.63 & 0.92 & 1.14 & 0.56 & 0.93 \\
$\rhoO$ & [g cm$^{-3}$] & $2.5 \times 10^{-9}$ & $2.4 \times 10^{-9}$ & $2.5 \times 10^{-9}$ & $2.9 \times 10^{-9}$ & $3.5 \times 10^{-9}$ & $4.3 \times 10^{-9}$\\
$\rhosnorm$ & & $3.2 \times 10^{-9}$ & $7.9 \times 10^{-4}$ & $2.6 \times 10^{-4}$ & $1.0 \times 10^{-9}$ & $1.0 \times 10^{-9}$ & $1.0 \times 10^{-9}$\\
$\epsilon$ & & $0.10$ & $0.10$ & $0.077$ & $0.073$ & $0.14$ & $0.12$\\
$\etaO$ & [cm$^2$ s$^{-1}$] & $5.4 \times 10^{14}$ & $4.5 \times 10^{14}$ & $2.0 \times 10^{14}$ & $1.4 \times 10^{14}$ & $7.6 \times 10^{14}$ & $2.5 \times 10^{14}$\\
$\etaH$ & [cm$^2$ s$^{-1}$] & $1.9 \times 10^{16}$ & $1.6 \times 10^{16}$ & $6.9 \times 10^{15}$ & $4.7 \times 10^{15}$ & $2.3 \times 10^{16}$ & $6.7 \times 10^{15}$\\
$\etaA$ & [cm$^2$ s$^{-1}$] & $1.6 \times 10^{15}$ & $1.4 \times 10^{15}$ & $5.3 \times 10^{14}$ & $3.2 \times 10^{14}$ & $1.7 \times 10^{15}$ & $4.0 \times 10^{14}$\\
$\ElsasserO \approx 1/\etaHnorm$ & & 1.6 & 2.0 & 6.6 & 12. & 1.2 & 6.8\\
$\zh$ & [au] & 0.0092 & 0.0087 & 0.0062 & 0.0045 & 0.0072 & 0.0037\\
$\zb$ & [au] & 0.031 & 0.032 & 0.055 & 0.11 & 0.034 & 0.086\\
$\zs$ & [au] & 0.38 & 0.13 & 0.085 & 0.15 & 0.29 & 0.13\\
$[B_{\rm r}/ B_{\rm z}]_{\rm b}$ & & 1.0 & 1.1 & 1.4 & 1.4 & 1.2 & 1.4 \\
$[B_{\rm \phi}/ B_{\rm z}]_{\rm b}$ & & -0.037 & -0.034 & -0.015 & -0.0090 & -0.049 & -0.015 \\
$\mdota$ & [$\msun$ yr$^{-1}$] & $1.0 \times 10^{-5}$ & $1.0 \times 10^{-5}$ & $1.0 \times 10^{-5}$ & $1.0 \times 10^{-5}$ & $1.6 \times 10^{-5}$ & $1.6 \times 10^{-5}$\\
$\mdotout/\mdotin$ & & \multicolumn{4}{c}{$1.3 \times 10^{-2}$} & \multicolumn{2}{c}{$3.5 \times 10^{-2}$}\\
\hline
\end{tabular}
\end{center}
\label{tab:fiducial}
\end{table*}

%----------------------------------------------
% GLOBAL WIND-DRIVING DISC SOLUTIONS
%---------------------------------------------- 
\section{1+1.5D wind-driving disc solutions} \label{sec:change_var}

Now that we have studied in detail the structure of two disc-wind models, we generalize our search to include a range of models with unique disc characteristics. We use the weak-wind model analysed in Section \ref{sec:comp:weak} as our fiducial model, and vary the parameters $\mdotin$, $\aO$, $\SigmaO$ and $p$ (equation \ref{eqn:surfdens2}) one by one to investigate the resulting effects on the wind-launching region. We specifically focus on how the radial extent and ejection/accretion ratio of the wind change with disc properties, since these may be compared to other observational and theoretical predictions for disc winds.

%---------- Impact of the mass accretion rate onto the star \mdotin ----------
\subsection{Impact of the mass accretion rate $\mdotin$} \label{sec:change_mdota}

We begin by exploring the result of modifying $\mdotin$, the accretion rate at the inner-most radius of the wind-launching region. This is important for studying how the wind would respond to variations in accretion rate over periods longer than a dynamical time $\tau_{\rm d}$ (i.e. the Keplerian orbital time), which could occur via fluctuations during quiescence, or throughout the restorative decline in accretion rate following an outburst event \citep[see][]{Audard:2014wu}. We define each 1+1.5D wind model by $\aO = 1.0$, and a surface density profile of $\Sigma(r) = 630 (r/$au$)^{-1}$ g cm$^{-2}$. 

\begin{figure}
	\centering
	\includegraphics[width=84mm]{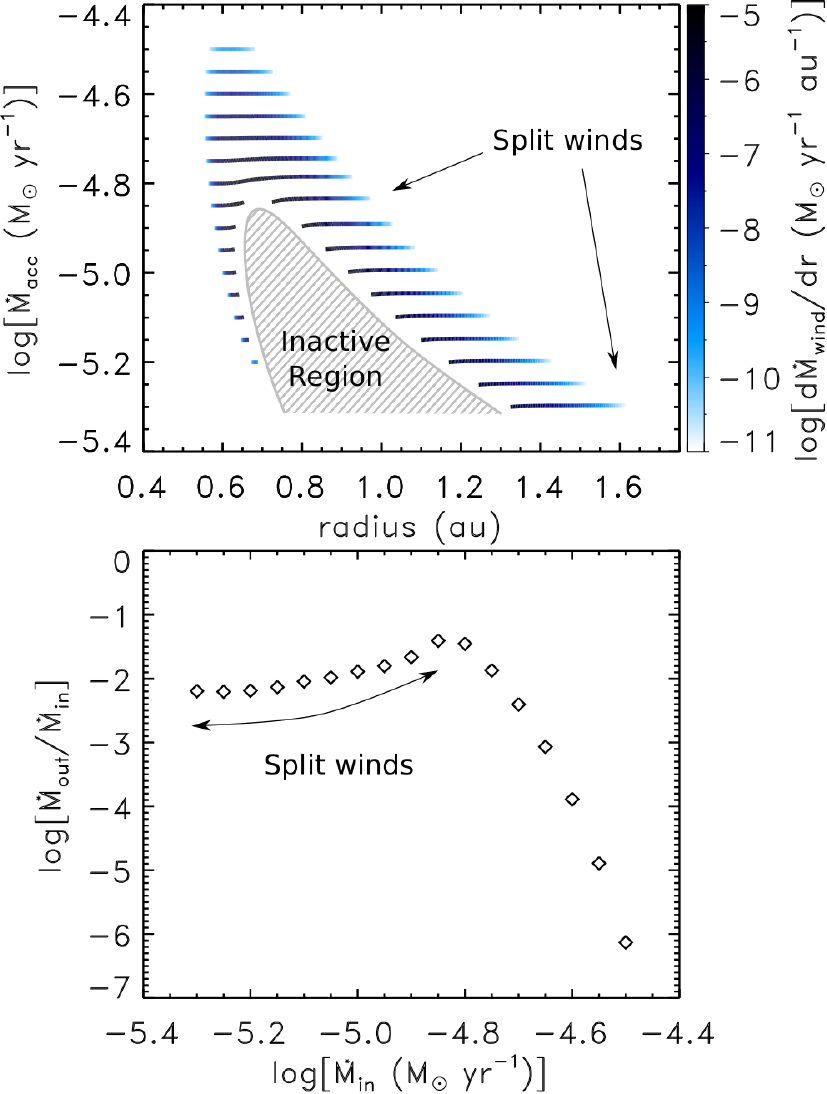}
	\caption{Local mass accretion rate $\mdota$ as a function of radius (top panel) and ejection/accretion ratio $\mdotout/\mdotin$ (bottom panel) for a series of 1+1.5D disc wind models with different $\mdotin$. Each disc is defined by an equipartition magnetic field strength at the midplane ($\aO = 1$), and a surface density profile \mbox{$\Sigma(r) = 630 (r/$au$)^{-1}$ g cm$^{-2}$}. The colouring in the top panel indicates the local wind mass loss rate at the disc surface along the radial extent of each model. The grey hatched area indicates the region where wind solutions are defined as unphysical due to the mass loading constraint; the `inactive' region (see Section \ref{sec:1.5D:const}).}
	\label{fig:change_mdota}
\end{figure}

The results of this investigation are shown in Fig. \ref{fig:change_mdota}. The top panel shows the local mass accretion rate as a function of radius, with each curve corresponding to a different choice of $\mdotin$. The wind-launching region for all models is radially localized, as can be observed from the colour-coded wind mass loss rates in the figure, with a maximum wind output between $r \sim 0.6$--$0.8$ au, and rapid attenuation on either side. For an explanation of why the wind is radially localized, see Section \ref{sec:comp:rates}.

As $\mdotin$ is decreased, the radial extent of the wind-launching region expands. This can be understood using the following explanation. At lower $\mdotin$, there is less material moving radially through the disc. This corresponds to less magnetic field bending in the radial and azimuthal directions, and hence a reduction of the magnetic compression of the disc. As we described in Section \ref{sec:comp:rates}, the less the disc is compressed, the more dense the resulting wind is. Hence lower values of $\mdotin$ lead to a higher wind mass loss rate across the radial extent of the wind, expanding the wind region where the mass loss rate is significant, and increasing the ejection/accretion ratio, as can be seen in the lower panel of Fig. \ref{fig:change_mdota}. 

Below $\mdotin = 10^{-4.8} \msun$ yr$^{-1}$, however, the wind-launching region is divided in two, resulting from an intermediate portion of the disc wind becoming ineffective for launching a wind. In this `inactive' region, the vertical mass flux would be so large that it would transport more angular momentum out of the disc than that brought in by the accretion flow, and similarly $\mdotw \nll \mdota$, leading to an unphysical launching configuration and no wind (see the mass loading constraint, Section \ref{sec:1.5D:const}). High mass loss rates in protostellar disc winds lead to instability \citep{Cao:2002cc}. As described above, lower values of $\mdotin$ lead to a higher wind mass loss rate across the radial extent of the wind, and hence a greater portion of the disc becomes magnetically inactive. As a result of the radial increase of the inactive region, $\mdotout/\mdotin$ steadily decreases.

%---------- Impact of the magnetic field strength parameter a_0 ----------
\subsection{Impact of the magnetic field strength parameter $\aO$} \label{sec:change_a0}

% Introduction
The strength of the magnetic field is important for determining not only the behaviour of disc winds, but also the characteristics of the MRI \citep[e.g.][]{Turner:2014tw}, and X winds \citep[e.g.][]{Shu:1994gr}. We now consider the impact of changing the magnetic field strength by calculating a series of 1+1.5D disc wind models with \mbox{$\mdotin = 1 \times 10^{-5} \msun$ yr$^{-1}$} and a surface density profile $\Sigma(r) = 630 (r/\mbox{au})^{-1}\mbox{ g cm}^{-2}$, while varying the midplane ratio of the Alfv\'{e}n speed to the sound speed, $\aO$ (see equation \ref{eqn:a0}).

\begin{figure}
	\centering
	\includegraphics[width=84mm]{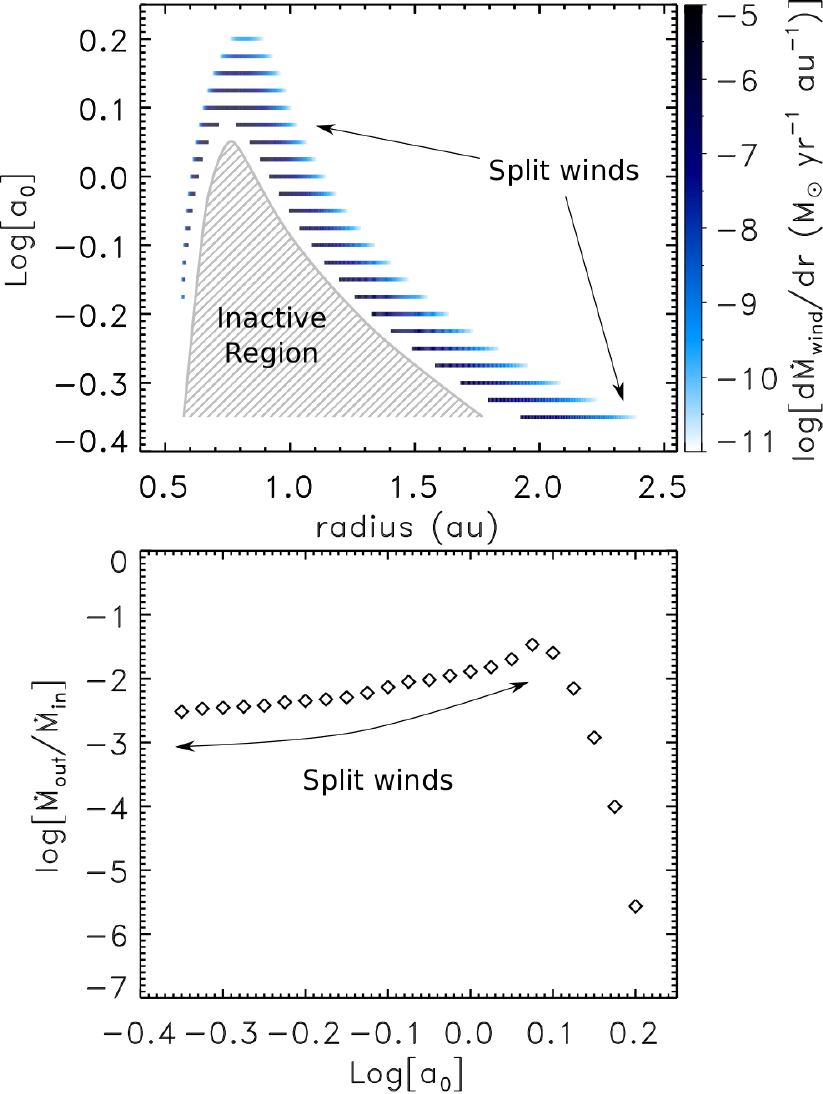}
	\caption{As per Fig. \ref{fig:change_mdota}, however this time varying the midplane magnetic field strength via $\aO \equiv \vAO/\cs$. For reference, the plasma beta is related to $\aO$ via the equation $\beta_{\rm 0} = 2/\aO^2$.}
	\label{fig:change_a0}
\end{figure}

%Radial variation
The results of varying $\aO$ are shown in Fig. \ref{fig:change_a0}, and display similar trends to those of the $\mdotin$ study. All solutions are radially localized, with maximum wind output around $r \sim 0.8$--$0.9$ au. As $\aO$ is decreased, the wind-launching region expands and the ejection/accretion ratio increases. This can be explained as follows. Similar to a decrease in $\mdotin$, if the magnetic field strength is decreased, the magnetically-compressed density scaleheight ($\zh$) increases due to less magnetic pressure, and the launch point ($\zb$) decreases, leading to a higher wind density, and hence larger wind output and ejection/accretion ratio. This is evidenced in the bottom panel of Fig. \ref{fig:change_a0}. Changing $\aO$ does significantly shift the radius where the wind output peaks, however, below $\aO = 1.25$ the launching region is divided in two due to a central section of no wind. As this inactive region expands at lower $\aO$, the split wind-launching regions move further away from the central radius. The splitting of the wind into two separate regions below $\aO = 1.25$ also disrupts the rapid increase in $\mdotout/\mdotin$ leading it into a steady decline.

%---------- Impact of the surface density normalization $\SigmaO$ ----------
\subsection{Impact of the surface density normalization $\SigmaO$} \label{sec:change_sigma0}

%Introduction
We now investigate how the behaviour of disc winds change with $\SigmaO$ (see equation \ref{eqn:surfdens2}). We define each model by \mbox{$\aO = 1.0$}, $\mdotin = 1 \times 10^{-5} \msun$ yr$^{-1}$, and a surface density profile \mbox{$\Sigma(r) = \SigmaO (r/\mbox{au})^{-1}$}, while varying the surface density coefficient $\SigmaO$. A change in $\SigmaO$ corresponds to a variation in the mass of the disc. We expect to observe this in discs that are being emptied out via quiescent mass accretion \citep[e.g.][]{Williams:2011js,Armitage:2015tz}, or on local scales as a result of episodic accretion \citep[see][]{Audard:2014wu}. Current observational estimates of $\SigmaO$ vary substantially in the range \mbox{$\SigmaO \sim 1$--$2000$ g cm$^{-2}$} \citep{Andrews:2007hv, Persson:2016dd}.

\begin{figure}
	\centering
	\includegraphics[width=84mm]{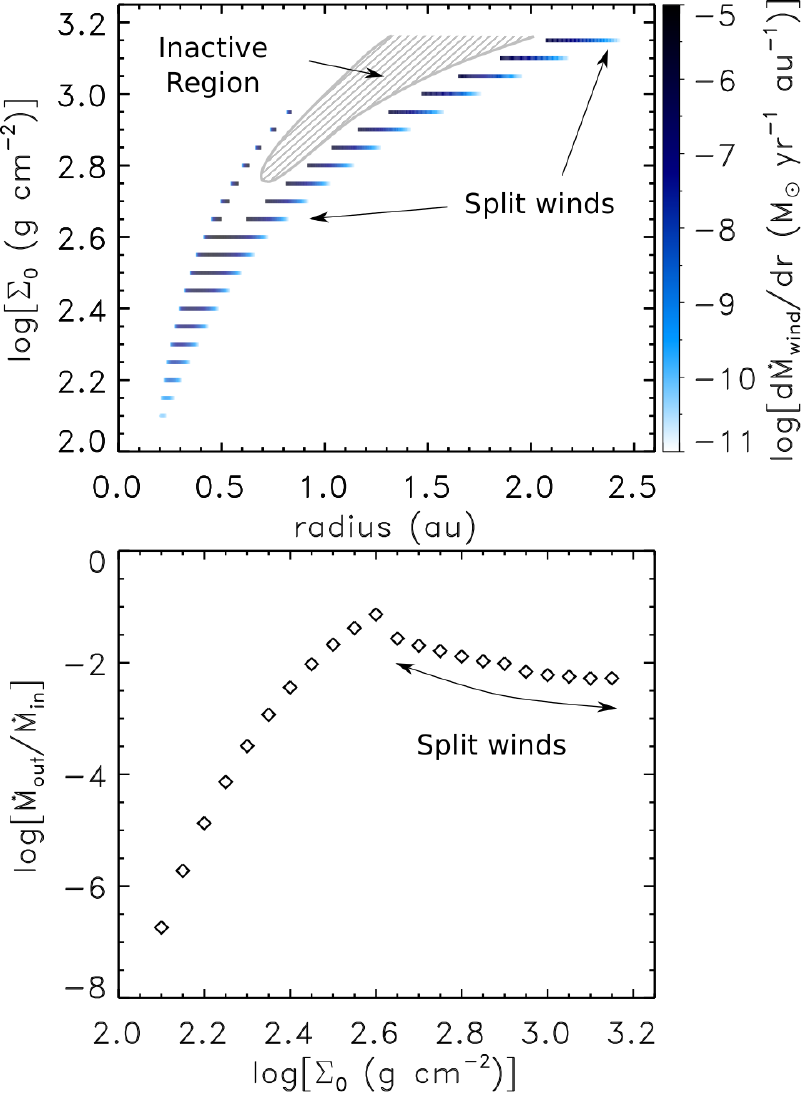}
	\caption{As per Fig. \ref{fig:change_mdota}, however this time varying the surface density profile, specifically $\SigmaO$ in the profile \mbox{$\Sigma(r) = \SigmaO (r/\mbox{au})^{-1}$}.}
	\label{fig:change_sigma0}
\end{figure}

%Radial extents
The results are given in Fig. \ref{fig:change_sigma0}. The top panel shows the variation in the radial extent of the wind as a function of $\SigmaO$. Similar to the previous two studies in Sections \ref{sec:change_mdota} and \ref{sec:change_a0}, all solutions are radially localized. For low $\SigmaO$, the launching region exists quite close to the protostar, and moves further out for larger values. Larger surface densities decrease the amount of radiation incident upon the disc midplane (see Appendix \ref{sec:ionization}), which leads to higher magnetic diffusivity and lower field-matter coupling (eqns. \ref{eqn:ohmdiff_ext}--\ref{eqn:ambidiff_ext} and \ref{eqn:elsasser}). Since the field-matter coupling determines the inner boundary of the wind-launching region (see Section \ref{sec:comp:rates} for details), and $\ElsasserO$ increases with radius (see Fig. \ref{fig:elsasser}), a lower overall $\ElsasserO$ means that the inner radius of the wind region moves outward, shifting the entire wind region to larger radii.

We also observe that the wind-launching region expands as $\SigmaO$ increases. This is a result of the power-law description of the surface density profile. As explained above, the wind-launching region is highly dependent on the value of the field-matter coupling, which is a function of surface density. Since the surface density profile is shallower at larger radii, we would expect that the optimal region for wind launching would have a larger extent. Since the wind has a larger extent at higher $\SigmaO$, it also has a larger ejection/accretion ratio, as seen in the bottom panel of Fig. \ref{fig:change_sigma0}. However, above $\SigmaO = 400 \mbox{ g cm}^{-2}$ the wind splits in two due to no wind being launched at intermediate radii, leading to a steady decrease in $\mdotout/\mdotin$.

%---------- Impact of the surface density exponent, p ----------
\subsection{Impact of the surface density exponent, $p$} \label{sec:change_exp}

\begin{figure}
	\centering
	\includegraphics[width=84mm]{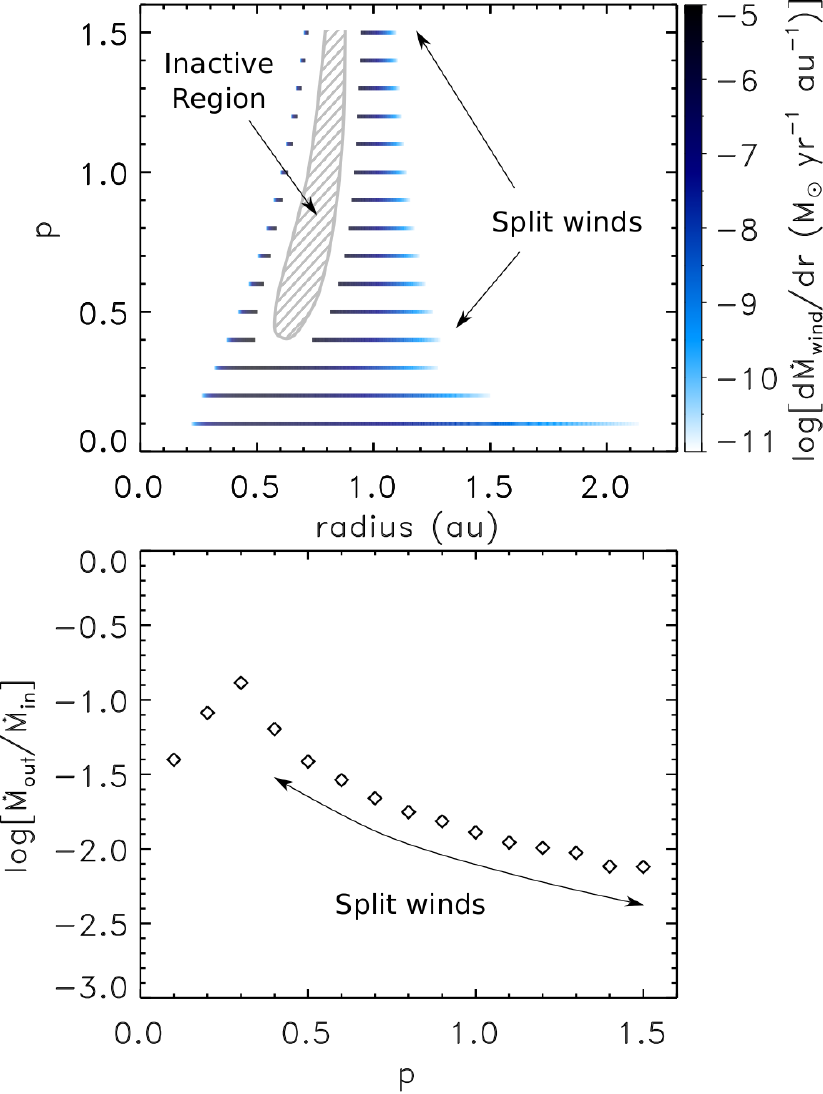}
	\caption{As per Fig. \ref{fig:change_mdota}, however this time varying the surface density profile, specifically $p$ in the profile \mbox{$\Sigma(r) = 630 (r/\mbox{au})^{-p}\mbox{ g cm}^{-2}$}.}
	\label{fig:change_p}
\end{figure}

Finally, we explore the impact of changing the power-law dependence of the surface density, defined by the parameter $p$ in equation (\ref{eqn:surfdens2}), on the properties of the wind-launching region. Current power-law estimates give values for $p$ most commonly between 0--1 \citep{Bergin:2011uma,Persson:2016dd}, while $p = 1.5$ in the MMSN model. We examine discs with $p$ between $0.1$ and $1.5$, assuming a disc characterized by $\aO = 1$, $\mdotin = 1 \times 10^{-5} \msun$ yr$^{-1}$, and a surface density profile $\Sigma(r) = 630 (r/\mbox{au})^{-p}\mbox{ g cm}^{-2}$.

The results are shown in Fig. \ref{fig:change_p}, with radial estimates for the extent of the disc wind as a function of $p$ displayed in the top panel. Similar to the other three studies of Section \ref{sec:change_var}, all solutions are radially localized, with no wind launching at intermediate radii for $p > 0.3$. For an almost flat profile ($p = 0.1$), the disc wind extends from $\sim 0.2$--2.1 au, and decreases in width as $p$ increases, until it is less than \mbox{0.4 au} wide for $p = 1.5$. This occurs because as $p$ increases, the surface density profile steepens, narrowing the region which provides field-matter coupling values favourable for wind launching.

Interestingly, $\mdotout/\mdotin$ decreases as $p$ is lowered (ignoring split winds). This occurs because for flatter surface density profiles, the maximum local wind output is lower than for steeper profiles. As seen in the other studies, once the wind-launching region splits in two, the ejection/accretion ratio steadily declines.

%----------------------------------------------
% DISCUSSION
%----------------------------------------------
\section{Discussion} \label{sec:discussion}

%Introduction
The results presented in Sections \ref{sec:comp} and \ref{sec:change_var} provide a detailed study into the properties of disc winds and how they vary with the characteristics of the underlying disc. We specifically focus on how the radial extent and ejection/accretion ratio of the winds change with disc properties, as these two attributes may be compared to observations and theoretical predictions for disc winds.

\subsection{Comparison with observations}
%Comment on radial extents.
For mass accretion rates in the range $10^{-5.4}$--$10^{-4.5} \msun$ yr$^{-1}$, midplane Alfv\'{e}n-to-sound speed ratios between $0.45$ and $1.6$, surface density profiles with $\SigmaO$ in the range $130$--$1400 \mbox{ g cm}^{-2}$ and $p$ between $0.1$ and $1.5$ (given \mbox{$\Sigma(r) = \SigmaO (r/\mbox{au})^{-p}$}), we obtain disc winds in the range of \mbox{0.2--2.4 au} from the central protostar. Observations vary on their predictions for where these winds exist, however a number of studies agree on a value between $\sim 0.3$--5 au \citep{Bacciotti:2002kr, Anderson:2003tp, Ray:2007tk}. On the other hand, theoretical predictions give values between 0.2 and 10 au \citep{Ferreira:2006vl, Pudritz:2007uu, Turner:2014tw}. Hence our numerical models are consistent with independent theoretical constraints and with observations. However, this does not mean that these are the only wind-driving regions in the disc. Other regions may exist for different relevant parameters.

We find that increasing the inner accretion rate $\mdotin$ while leaving the midplane ratio of the Alfv\'{e}n speed to the sound speed ($\aO$) and the surface density profile unchanged, moves the disc wind region inwards. Similarly, increasing $\aO$ independent of the other parameters has the same effect. On the other hand, increasing the overall surface density via $\SigmaO$ (see equation \ref{eqn:surfdens2}) moves the disc wind outwards. Decreasing the power-law index of the surface density $p$ widens the extent of the disc wind.

%Comment on mdotout/mdotin ranges.
We also find that the ejection/accretion ratios \citep[$\mdotout/\mdotin$, see][]{Cabrit:2007hz} of our disc wind models are in the range $10^{-1}$--$10^{-7}$ for the parameter combinations investigated here. We regard these values to be upper limits for realistic wind-launching regions, given our assumptions for the mass loading constraint derived in Section \ref{sec:1.5D:const}. For this constraint we assume that the angular momentum contained by the magnetic field is transferred back into the flow on a length scale $z \approx r$ above the disc midplane. In practice this could occur at even greater heights, which would increase the radial range of the intermediate magnetically-inactive region and hence decrease the ejection/accretion ratios in our models. The range of $\mdotout/\mdotin$ we find agrees closely with the 1.5D predictions of WK93 and \citet{Pelletier:1992fk}, who calculate values for $\mdotout/\mdotin$ between $10^{-1}$ and $10^{-5}$. Observational estimates of $\mdotout/\mdotin$ also overlap our results, with values between $10^{-1}$ and $10^{-4}$ \citep{Cabrit:2007hz,Ellerbroek:2013bf,Watson:2015uo}. From our findings, $\mdotout/\mdotin$ generally increases as $\mdotin$ and $\aO$ decrease, while the opposite is true for $\SigmaO$ and the power-law index $p$. However, each of these trends is altered significantly by excessive mass loading of the field lines at intermediate radii, causing the fluid configuration in this region to be ineffective for wind launching (see Section \ref{sec:change_mdota} and Fig. \ref{fig:change_mdota}). This leads to a steady decrease in $\mdotout/\mdotin$ with each parameter as the magnetically inactive region widens.

\subsection{Comparison with current global simulations}

Many studies have been performed using the shearing box approximation to investigate disc structure and wind launching in protostellar discs \citep[e.g.][]{Suzuki:2009wl, Suzuki:2010ul, Fromang:2013vf, Bai:2013ue, Bai:2013bx, Lesur:2013ha, Simon:2013fw, Bai:2014ea}. Compared to the 1+1.5D model used in this paper, the shearing box approximation has a few advantages. Due to its 3D nature, the shearing box approximation is able to model turbulence and other time-dependent properties such as disc evolution and chemical mixing much more thoroughly than our 1+1.5D method. However, when it comes to radial properties and outflow rates, the shearing box approximation is severely limited by its boundary conditions \citep[see][]{Turner:2014tw}. Hence, while both types of models have unique advantages, quasi-global 2.5 or 3D time-dependent models such as \citet{Gressel:2015ks}, with higher resolution are needed to make progress.

Comparing our results to the quasi-global model of \citet{Gressel:2015ks}, we find quite different results between the two. Their model is time-dependent, includes ambipolar and Ohm diffusivities, and is initialized with a weak magnetic field ($\beta_{\rm 0} = 10^{\sim 5 - 7}$), while our model is steady-state, includes Hall diffusion as well as ambipolar and Ohm, and maintains a much stronger magnetic field ($\beta_{\rm 0} = 10^{\sim 0 - 1}$). 

The weak magnetic field configuration of \citet{Gressel:2015ks} leads to a laminar region about the disc midplane, with minimal accretion and field-line bending . This is in stark contrast to our model, which displays much higher inner-disc accretion rates and field-line bending. Given that the study by \citet{Moll:2012co} employs magnetic field strengths much closer to ours, resulting in a similar field configuration, we expect that the morphological difference between our model and that of \citet{Gressel:2015ks} is related to the magnetic field strength threading the disc \citep[see also the weakly coupled models of][]{Wardle:1997vy}.

With regard to the wind-launching region, \citet{Gressel:2015ks} find winds launched continuously along all radii within their simulation domain (0.5 - 5.5 au), while the winds produced by our models are limited to a narrow range, often marked with radial gaps where no wind is launched. This could again be related to the magnetic field strength within the disc, therefore, it would be beneficial for future global studies to focus on the intermediate plasma beta range ($\beta_{\rm 0} = 10^{\sim 1 - 4}$), to observe how the field configuration, accretion and outflow rates change between the weak regime, which is optimal for MRI growth, and the strong regime, which is optimal for centrifugal wind launching. 

\subsection{Implications}

%Comment on the profile of dMdota/dr
The significant results from this study are that disc winds tend to be radially localized, meaning that the majority of the wind output is centralized around a particular radius, and the discovery of a new class of disc winds containing an ineffective launching configuration at intermediate radii. The localization of wind output reinforces the applicability of deriving the launch radius of protostellar jets based on their poloidal velocities \citep[e.g.][]{Ferreira:2006vl, AgraAmboage:2011tg, White:2014cj}, while an ineffective or unstable launching region at intermediate radii could contribute to the knots we see in protostellar jets \citep[e.g.][]{Frank:2014wl}

%Temporal implications of mdotin
Variation of key disc parameters could represent the changing structure of the disc with time. For example, $\mdotin$ could change as a result of fluctuations during quiescent intervals, or steadily decline following an outburst event \citep{Audard:2014wu}. In the case of a declining accretion rate, we expect the wind-launching region to shift to larger radii and the $\mdotout/\mdotin$ to increase, or in the case of a split wind, to decrease (see Fig. \ref{fig:change_mdota}). 
Similarly, we expect $\SigmaO$ to decrease over time due to mass accretion onto the central protostar \citep{Williams:2011js,Armitage:2015tz}, or vary on local scales as a result of accretion outbursts \citep{Audard:2014wu}. From the results in Section \ref{sec:change_sigma0}, the disc wind would move inward as the disc mass is depleted, and $\mdotout/\mdotin$ would change accordingly.

\subsection{Model limitations} \label{sec:limitations}

%Mass accretion rates and Lambda constant with height.
We find that the radial extents and ejection/accretion ratios of the models presented in this paper are in good agreement with observations for our choice of disc parameters. The range of $\mdotin$ found to launch viable disc winds is relatively high compared with those observed in Classical (Class II) T-Tauri discs \citep[$\sim 10^{-8}$ $\msun$ yr$^{-1}$,][]{Shariff:2009hj}. These accretion rates overlap those inferred for FU Orionis objects (FUors, $10^{-6}$--$10^{-3}$ $\msun$ yr$^{-1}$) and EX Lupi objects (EXors, \mbox{$10^{-8}$--$10^{-6}$ $\msun$ yr$^{-1}$}), classes of pre-main-sequence stars which exhibit episodic accretion over time-scales of several decades, or years, respectively \citep{Aspin:2010hh,Audard:2014wu}. Hence our models may be representative of discs in an accretion outburst phase. However, it is also likely that the assumption of constant field-matter coupling $\Lambda$ with height, or equipartition magnetic fields ($\aO \lesssim 1$), are a major contributor to the large accretion rates in our models. Discs with variable $\Lambda$ and weaker coupling exhibit a markedly different structure, with lower accretion rates due to an inner magnetically dead zone \citep[see][]{Wardle:1997vy}. If the constant-$\Lambda$ condition is relaxed, $\Lambda$ is expected to initially increase with height above the disc midplane as the column density (which shields the disc from ionizing radiation) diminishes, leading to larger ionization fractions. However, above a certain height this effect is countered by a rapid decrease in density, leading to a reduction in the field-matter coupling \citep[see figure 7.5 of][]{Konigl:2011tk}. This change in $\Lambda$ with height can potentially lead to conditions at the surface of the disc which differ from the models presented in this study, and hence different wind properties. This is therefore the next logical step in our study, and will be addressed in a subsequent paper.

%Limitation: Constant $\aO$.
In addition to the constant-$\Lambda$ approximation, all results contained in this paper assume that the magnetic field parameter $\aO$ (the ratio of the Alfv\'{e}n speed to the sound speed at the disc midplane) does not change with radius. Assuming that the Alfv\'{e}n speed scales as the Keplerian velocity, BP82 showed that $\aO$ is constant with radius for an ideal MHD self-similar disc. Therefore, this may be considered as an intuitive first approximation. We expect however, that in a realistic disc, $\aO$ would vary with radius. This could in theory be constrained by systematically conserving the vertically integrated angular momentum flux, in a similar approach to our treatment of the mass flux (see Section \ref{sec:2d}), however we have neglected this approach for the time being due to the added complexity.

%No drifting field lines $\epsilonB = 0$. 
We also neglect the radial drift of poloidal magnetic field lines (i.e. we assume that $\epsilonB = 0$) in each of the 1.5D solutions that make up the 1+1.5D models. Employing a similar radially-localized model to the one used here, but specialized to the ambipolar diffusion limit, WK93 derived solutions for positive and negative values of $\epsilonB$ and found that solutions with the same value of ($\epsilon - \epsilonB$) are qualitatively similar. We use this result to justify the selection of $\epsilonB = 0$ in our models, as this should not significantly impact the generality of the results. For a full discussion of the $\epsilonB = 0$ approximation, see Appendix A of KSW10.

The disk is also assumed to be vertically isothermal, and follows a radial profile as given by the MMSN model. The temperature may actually increase towards the surface, as a result of thermal decoupling of dust and gas, due to the low density. This would affect the conditions at the sonic point, and mass flux.

%Limitation: No grains.
Finally, we assume that the charged species are exclusively ions and electrons. This is a reasonably valid approximation at late times in the disc's evolution when dust grains have settled to the midplane, and hence grains may be neglected when considering the disc structure at larger $z$ \citep[e.g.][KSW10]{Dullemond:2004jj}. Other accretion mechanisms, such as turbulence generated by the MRI, could in principle stir up the dust component to larger heights above the midplane, however, in the radial locations of the disc we explore in this paper ($\sim 0.2$--2.4 au) the MRI is thought to have little to no influence due to turbulent quenching via the Hall effect \citep{Turner:2014tw}. This being said, due to the flexibility of the conductivity tensor formulation in each of our models, we could in principle extend our results to include the influence of dust grains on disc wind morphology \citep{Wardle:1999um}.

%----------------------------------------------
% CONCLUSIONS
%----------------------------------------------
\section{Summary and Conclusions} \label{sec:conclusions}

%Introduction
In this paper we investigated the properties of disc winds and how they vary with the characteristics of the underlying disc. Using the first 1+1.5D steady-state disc wind model incorporating all three diffusion mechanisms (Ohm, Hall and ambipolar) and the effects of X-rays, cosmic rays and radioactive decay, we determined how the radial extent and ejection/accretion ratio ($\mdotout/\mdotin$) of the wind-launching region varies with accretion rate $\mdotin$, magnetic field strength (parametrized by $\aO \equiv \vAO/\cs$) and surface density profile given by $\Sigma(r) = \SigmaO (r/$au$)^{-p}$. Each of these parameters has a significant effect on the radial position and extent of the wind-launching region, and the ejection/accretion ratio $\mdotout/\mdotin$, while still confining them to within observational and theoretical estimates. In summary, we found that:
\begin{itemize}
\item All 1+1.5D wind solutions are radially localized (i.e. the wind mass flux peaks at a particular radius, and rapidly drops off on either side). At smaller radii, the wind attenuates as a result of decreased field-matter coupling, while at larger radii, magnetic compression of the disc combined with a higher disc surface results in a lower density wind and decreasing outflow rate.
\item Many 1+1.5D wind solutions are split into two parts by an ineffective launching configuration at intermediate radii, where no wind is expected to exist due to excessive mass loading of the field lines. This inactive region has a substantial impact on the ejection/accretion ratio, leading to a lower $\mdotout/\mdotin$. More detailed simulations are required to determine the behaviour of the disc and wind in this region.
\item Decreasing $\mdotin$ expands the radial extent of the wind-launching region while moving it to larger radii, and increases $\mdotout/\mdotin$ (ignoring the effects of the inactive region). This is a result of less magnetic field bending in the radial and azimuthal directions, and hence reduced magnetic compression of the disc.
\item Similarly, lower values of $\aO$ lead to an expanded disc due to less magnetic pressure, extending the wind-launching region in both directions, and increasing $\mdotout/\mdotin$.
\item Decreasing the surface density power-law constant $\SigmaO$ has the opposite effect. At lower $\SigmaO$ (representative of a lower mass disc), the disc wind is launched closer to the protostar and has a reduced radial extent as a result of the modification of the field-matter coupling profile. This reduction in the radial extent leads to lower values for $\mdotout/\mdotin$.
\item Decreasing the surface density power-law index $p$ has the effect of radially stretching the launching region, as the region of optimal field-neutral coupling in the disc widens. It also decreases $\mdotout/\mdotin$ in the process.
\end{itemize}
In conclusion, we find that changes in the physical properties of protostellar discs have an important impact on the position of the launching region and the power of disc winds. We do stress though, that real protostellar discs have a much more complex morphology, including stratified diffusion regimes, regions of magnetohydrodynamic turbulence, and dust grain populations which we have not taken into account. Thus, our study provides a first look at the 1+1.5D structure of disc winds, while more detailed investigations are required to make future progress.

%----------------------------------------------
% ACKNOWLEDGEMENTS
%---------------------------------------------- 
\section*{Acknowledgements}
We thank the anonymous referee for their thorough and constructive report, which improved the paper significantly. We also thank Mark Wardle for useful discussions. CAN acknowledges support by an Australian Postgraduate Award. CF gratefully acknowledges funding by the Australian Research Council's Discovery Projects (grant~DP150104329 and DP170100603). CF thanks for high performance computing resources provided by the Leibniz Rechenzentrum and the Gauss Centre for Supercomputing (grants~pr32lo, pr48pi and GCS Large-scale project~10391), the Partnership for Advanced Computing in Europe (PRACE grant pr89mu), the Australian National Computational Infrastructure (grant~ek9), and the Pawsey Supercomputing Centre with funding from the Australian Government and the Government of Western Australia, in the framework of the National Computational Merit Allocation Scheme and the ANU Allocation Scheme. This research was supported by ARC grant No. DP120101792.

%%%%%%%%%%%%%%%%%%%%%%%%%%%%%%%%%%%%%%%%%%%%%%%%%%



\bibliographystyle{mnras}
\bibliography{bibliography.bib}

%%%%%%%%%%%%%%%%%%%%%%%%%%%%%%%%%%%%%%%%%%%%%%%%%%

%%%%%%%%%%%%%%%%% APPENDICES %%%%%%%%%%%%%%%%%%%%%

\appendix

\section{Resolution study}
\label{sec:res}

Here we present the results of a resolution study undertaken to determine the convergent resolution for the 1+1.5D models. We perform a series of identical simulations, using the outer region of the weak-wind model in Section \ref{sec:comp:weak}, at increasing resolutions until the ejection/accretion ratio and radial range of the wind-launching region converge. These models are characterized by $\aO = 1$, $\mdotin = 10^{-5} \msun$ yr$^{-1}$ and a surface density profile of $\Sigma(r) = 630 (r/{\rm au})^{-1}$ g cm$^{-2}$. The results of the study are shown in Table \ref{tab:res}. From the table we see that the simulations are converged for $k \gtrsim 1000$, where $k$ is the number of 1.5D solutions per decade of radius (see equation \ref{eqn:k}). Hence we use a resolution of $k = 1000$ for all 1+1.5D models presented in this paper.

\begin{table}
\caption{Selected properties of a representative 1+1.5D model (the outer region of the weak-wind model in Section \ref{sec:comp:weak}), characterized by $\aO = 1$, $\mdotin = 10^{-5} \msun$ yr$^{-1}$, and $\Sigma(r) = 630 (r/{\rm au})^{-1}$ g cm$^{-2}$, at increasing resolutions. From left to right, these properties are the resolution per decade of radius, the number of solutions spanning the wind-launching region, the ejection/accretion ratio, and the radial range of the wind. Models with $k \gtrsim 1000$ are representative of the converged system.}
\begin{center}
\begin{tabular}{clcl}
\hline
$k$ & N$_{\rm solutions}$ & $\mdotout/\mdotin$ & r$_{\rm range}$ (au) \\
\hline
20 & 2 & $2.140 \times 10^{-3}$ & 0.1220 \\
50 & 4 & $5.145 \times 10^{-3}$ & 0.1415 \\
100 & 9 & $7.391 \times 10^{-3}$ & 0.1888 \\
200 & 19 & $8.528 \times 10^{-3}$ & 0.2124 \\
500 & 49 & $9.042 \times 10^{-3}$ & 0.2267 \\
1000 & 97 & $8.698 \times 10^{-3}$ & 0.2267 \\
2000 & 194 & $8.795 \times 10^{-3}$ & 0.2277 \\
4000 & 388 & $8.818 \times 10^{-3}$ & 0.2282 \\
\hline
\end{tabular}
\end{center}
\label{tab:res}
\end{table}

\section{Div B study}
\label{sec:divb}

\begin{figure}
	\centering
	\includegraphics[width=84mm]{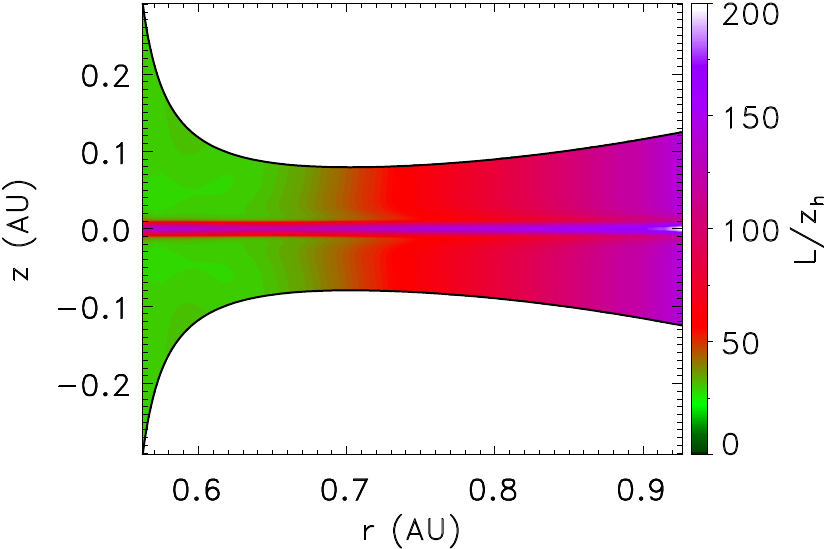}
	\caption{The local ratio of $L/\zh$ in the $(r,z)$-plane, for the strong-wind model described in Section \ref{sec:comp:weak}, where $L$ is the vertical scale length of $B_{\rm z}$, derived using $B_{\rm r}$ and $\nabla \cdot \mathbf{B} = 0$ (see equation \ref{eqn:l}), and $\zh$ is the disc scale height.}
	\label{fig:divb}
\end{figure}

We can prove that the 1+1.5D models satisfy the $\nabla \cdot \mathbf{B} = 0$ constraint by comparing the estimated local scale length $L$ for $B_{\rm z}$ (derived using $B_{\rm r}$ and $\nabla \cdot \mathbf{B} = 0$; see equation \ref{eqn:l}), with the disc scale height $\zh$. The method for calculating $L$ is outlined in Section \ref{sec:2d:const}. Fig. \ref{fig:divb} displays the local ratio of $L/\zh$ in the $(r,z)$-plane for the strong wind model described in Section \ref{sec:comp:weak}. From the figure we see that within the disc, the minimum value for this ratio is 25, and increases with radius and to infinity at the disc midplane. Hence we can confirm beyond doubt that the $\nabla \cdot \mathbf{B} = 0$ condition is satisfied to within error in our 1+1.5D models.

\section{Ionization model}

\label{sec:ionization}

In order to calculate the three diffusivity components (equations \ref{eqn:ohmdiff_ext} - \ref{eqn:ambidiff_ext}) and close the system of non-ideal equations, we use the following model to calculate the electron density at the midplane. We adopt the simplification of KSW10, i.e. that the ratios of the diffusivities are constant with $z$. These ratios are calculated at the midplane and then applied to the vertical structure of the disc. We assume that grains have settled out and that the electron (and ion) density is determined by the equilibrium ionization balance equation \citep[e.g.][equation 35]{Salmeron:2005dg}, in the limit where the dominant recombination mechanism is the radiative recombination of metal ions \citep{Fromang:2002iw, Salmeron:2003by}. The electron density is then given by 
\begin{equation}
n_{\rm e} (r) = \left( \frac{\zeta n_{\rm H}}{\alpha} \right)^{\frac{1}{2}}, \label{eqn:n_e}
\end{equation}
where $\alpha = 3 \times 10^{-11} T^{-1/2}$ cm$^{3}$ s$^{-1}$ is the radiative recombination rate for the metal ions \citep{Salmeron:2005dg} and $n_{\rm H} = \rho/1.4 m_{\rm H}$. The ionization rate $\zeta$ is made up of contributions from mechanisms which may be active at the disc midplane, namely cosmic rays ($\zeta_{\rm CR}$), X-rays ($\zeta_{\rm XR}$) and radioactivity of nuclides within the disc ($\zeta_{\rm R}$), therefore $\zeta = \zeta_{\rm CR} + \zeta_{\rm XR} + \zeta_{\rm R}$. Thermal ionization is not considered in our models, as it is most relevant for temperatures $> 1000$ K, or for $r \lesssim 0.1$ au assuming the minimum mass solar nebula (MMSN) radial temperature profile of \citet{Hayashi:1981ws} (see equation \ref{eqn:temp}). We do not incorporate a vertical temperature profile and also ignore the high-temperature atmosphere likely to be present above the midplane \citep[see][]{Glassgold:2004hs}. We also exclude ultraviolet photons, as they have penetration depths $< 0.1$ g cm$^{-2}$ \citep{PerezBecker:2011dv} and are relevant in the surface layers only.
 
We calculate the ionization rate contributed by cosmic rays according to the model outlined in \citet{Salmeron:2005dg}, as it applies to the disc midplane
\begin{equation}
\zeta_{\rm CR} = 10^{-17} {\rm \exp} (- \Sigma/2 \lambda_{\rm CR})\mbox{ s}^{-1}, \label{eqn:cosmicrays}
\end{equation}
where $\Sigma$ is the disc surface density and $\lambda_{\rm CR} = 96$ g cm$^{-2}$ is the attenuation length for cosmic ray penetration \citep{Umebayashi:1981ve}.

Stellar X-rays can penetrate a column density \mbox{$\sim$10 g cm$^{-2}$} \citep{Igea:1999ei}, and can therefore contribute to the ionization balance at the disc midplane only at large radii and low surface densities. We estimate the ionization rate contributed by this agent by using the following fit by \citet{Bai:2009er} to the Monte Carlo radiative transfer results of \citet{Igea:1999ei}, assuming a dominant X-ray photon energy of $T_{\rm X} = 3$ keV,
\begin{eqnarray}
\zeta_{\rm XR} &= &\frac{L_{\rm X}}{10^{29} \mbox{ erg s}^{-1}} \left(\frac{r}{1 \mbox{ au}} \right)^{-2.2} \nonumber \\
&& {}\times (2 \zeta_{\rm 1} {\rm \exp}^{-(N_{\rm H}/N_{\rm 1})^{0.4}} \nonumber \\
&& {}+ 2 \zeta_{\rm 2} {\rm \exp}^{-(N_{\rm H}/N_{\rm 2})^{0.65}} ) \mbox{ s}^{-1}. \label{eqn:xrays}
\end{eqnarray}
In this fit, the intensity of direct X-rays is \mbox{$\zeta_{\rm 1} = 6 \times 10^{-12}$ s$^{-1}$} and that of scattered X-rays is $\zeta_{\rm 2} = 10^{-15}$ s$^{-1}$, with respective penetration columns of $N_{\rm 1} = 1.5 \times 10^{21}$ cm$^{-2}$ and \mbox{$N_{\rm 2} = 7 \times 10^{23}$ cm$^{-2}$} \citep{Bai:2009er}. The column density of hydrogen nuclei at the midplane $N_{\rm H}$ is related to the disc surface density via $N_{\rm H} = (\Sigma/2)/(1.4 m_{\rm H})$ and we assume the X-ray luminosity of the protostar to be \mbox{$L_{\rm X} = 10^{29}$ erg s$^{-1}$}.

Finally, we estimate the ionization rate associated with the decay of radioactive elements present in the disc (mainly $^{40} K$) in terms of the fraction of heavy metal elements in the gas phase $\delta_{\rm 2}$, and the abundance of grains relative to that of molecular clouds $f_{\rm g}$:
\begin{equation}
\zeta_{\rm R} = 6.9 \times 10^{-23} \left[\delta_{\rm 2} + (1 - \delta_{\rm 2}) f_{\rm g} \right] \mbox{ s}^{-1} \label{eqn:radioactivity}
\end{equation}
\citep{Umebayashi:1981ve, Umebayashi:1990vw}, where $\delta_{\rm 2} = 0.02$. The parameter $f_{\rm g}$ is set to zero \citep{Sano:2000wo} in accordance with the assumption that dust grains have settled to the disc midplane.

\begin{figure}
	\centering
	\includegraphics[width=84mm]{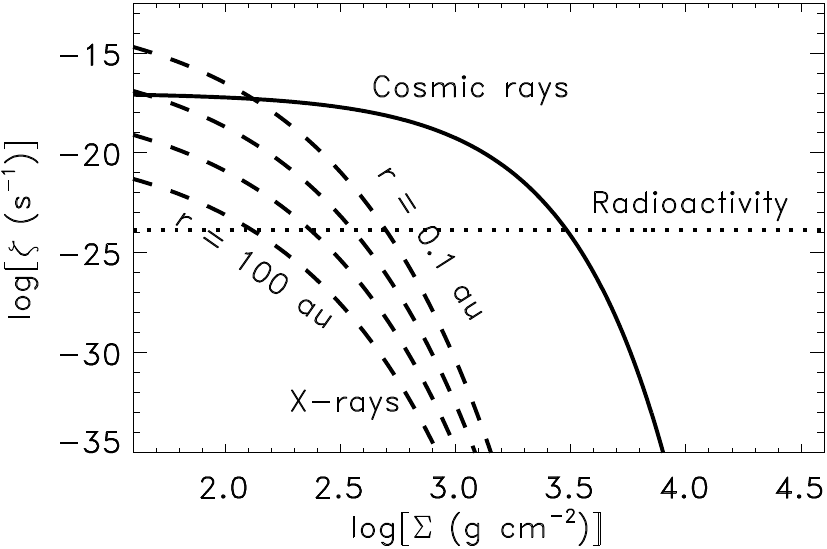}
	\caption{Ionization rates [s$^{-1}$] at the disc midplane contributed by cosmic rays (solid line, equation \ref{eqn:cosmicrays}), X-rays (dashed lines, equation \ref{eqn:xrays}) and radioactive decay (dotted line, equation \ref{eqn:radioactivity}), as a function of the surface density. The X-ray ionization rates are displayed for $r = 0.1$, 1, 10 and 100 au.}
	\label{fig:ionization}
\end{figure}

The contributions of each of the three ionization mechanisms to the total ionization rate at the disc midplane are shown in Fig. \ref{fig:ionization} for comparative purposes.

%%%%%%%%%%%%%%%%%%%%%%%%%%%%%%%%%%%%%%%%%%%%%%%%%%

% Don't change these lines
\bsp	% typesetting comment
\label{lastpage}
\end{document}